\documentclass[journal,10pt]{IEEEtran}

\usepackage{amssymb}
\usepackage{amsmath}
\usepackage{cite}
\usepackage{url}
\usepackage{xcolor}
\usepackage{cite,graphicx,amsmath,amssymb}
\usepackage{subfigure}
\usepackage{citesort}
\usepackage{fancyhdr}
\usepackage{mdwmath}
\usepackage{mdwtab}
\usepackage{caption}
\usepackage{amsthm}
\usepackage{amsmath}
\usepackage{bbm}
\usepackage{epstopdf}

\newtheorem{remark}{Remark}
\newtheorem{theorem}{Theorem}

\newtheorem{lemma}{Lemma}

\newtheorem{corollary}{Corollary}

\newtheorem{proposition}{Proposition}

\makeatletter
\def\ScaleIfNeeded{%
\ifdim\Gin@nat@width>\linewidth \linewidth \else \Gin@nat@width
\fi } \makeatother

\begin{document}

\title{Semi-Grant-Free NOMA: A Stochastic Geometry Model}


\author{

Chao~Zhang,
Yuanwei~Liu,~\IEEEmembership{Senior Member,~IEEE},
Zhijin~Qin,~\IEEEmembership{Member,~IEEE} and
Zhiguo~Ding,~\IEEEmembership{Fellow,~IEEE}

\thanks{C. Zhang, Y. Liu and Z. Qin are with the School of Electronic Engineering and Computer Science, Queen Mary University of London, London, UK (email:\{chao.zhang, yuanwei.liu, z.qin\}@qmul.ac.uk).}
\thanks{Z. Ding is with the School of Electrical and Electronic Engineering, The University of Manchester, Manchester, UK (e-mail: zhiguo.ding@manchester.ac.uk).}
}

\maketitle
\begin{abstract}
Grant-free (GF) transmission holds promise in terms of low latency communication by directly transmitting messages without waiting for any permissions. However, collision situations may frequently happen when limited spectrum is occupied by numerous GF users. The non-orthogonal multiple access (NOMA) technique can be a promising solution to achieve massive connectivity and fewer collisions for GF transmission by multiplexing users in power domain. We utilize a semi-grant-free (semi-GF) NOMA scheme for enhancing network connectivity and spectral efficiency by enabling grant-based (GB) and GF users to share the same spectrum resources. With the aid of semi-GF protocols, uplink NOMA networks are investigated by invoking stochastic geometry techniques. We propose a novel \textit{dynamic protocol} to interpret which part of the GF users are paired in NOMA transmissions via transmitting various channel quality thresholds by an added handshake. We utilize open-loop protocol with a fixed average threshold as the benchmark to investigate performance improvement. It is observed that dynamic protocol provides more accurate channel quality thresholds than open-loop protocol, thereby the interference from the GF users is reduced to a large extent. We analyze the outage performance and diversity gains under two protocols. Numerical results demonstrate that dynamic protocol is capable of enhancing the outage performance than open-loop protocol.
\end{abstract}

\begin{IEEEkeywords}
{D}ynamic protocol, grant-free, grant-based, stochastic geometry, uplink NOMA
\end{IEEEkeywords}

\section{Introduction}

The past decades have witnessed a huge proliferation of devices or services, i.e., 1) mobile devices such as sensors, machines and robots and 2) mobile services such as mobile online videos and mobile pay, which has led to an explosive growth of mobile broadband traffic \cite{IoT1,IoT2}. Aiming at enhancing the user experience of the proliferating devices, three requirements of the road map of the fifth-generation (5G) have been proposed, namely enhanced mobile broadband (eMBB), ultra-reliable low latency communications (URLLC) and massive machine-type communications (mMTC) \cite{JA5G,RWH5G}. Different from the aim of eMBB, i.e. to achieve high capacity and fast data rate with high energy efficiency, URLLC as a novel requirement differentiates from the others, which focuses on the trade-off between low latency and satisfying reliability. To facilitate the achievement of URLLC, the grant-free (GF) transmission scheme as a promising paradigm is widely utilized in the internet of things (IoT) networks to obtain low latency uplink transmission. In a nutshell, the concept of the GF transmission scheme can be regarded as the traditional grant-based (GB) transmission scheme by removing uplink scheduling requests (SR) and dynamic scheduling grants (SG) \cite{st26,LZ}. For clarification, compared with traditional GB random access schemes, the GF users are admitted to transmitting messages whenever they have data to send without any permission from the base station. Hence, the time consumption by lengthy handshakes is economized by achieving a low latency transmission scheme \cite{Kyang}. However, a significant shortcoming, frequent collision situations, causes the inability of multi-user detection, which is the most essential challenge for the GF transmission strategy \cite{st27,st28}. Therefore, non-orthogonal multiple access (NOMA) philosophy can be the promising solution of the coexistence between low latency and few collisions.

The explosively increasing devices of IoT networks and mobile internet pose the challenges of URLLC in 5G such as the trade-off between low latency and high reliability under the GF transmissions scheme \cite{5G1}. To cope with the aforementioned challenges above, NOMA can be universally utilized as a paradigm \cite{5G2}. More specifically, thanks to code-domain or power-domain multiplexing schemes and successive interference cancellation (SIC) \cite{SIC} technologies, multiple devices served by NOMA philosophy enable devices to share the same channel resources in a time block with few errors \cite{yuanwei5GNOMA,SM}. Thus, although the GF users may frequently collide in the same time block, messages can be successfully decoded via various power levels or codebooks by NOMA technology, which effectively solves the collision problems. Thus, limited spectrum resources can be shared by multiple users with low latency and low signaling overhead under the GF-NOMA transmission scheme \cite{zhiguo5G4,codeNOMA,17}.

\vspace{-1mm}

\subsection{Related Works and Motivation}

Since the design of GF transmission scheme aims at low latency by the cancelation of uplink grants, we investigate uplink GF NOMA designs to achieve high reliability. Extensive research contributions have explored the potential performance enhancement brought by uplink NOMA scheme as benchmarks. Typical models of multiple access designs in uplink NOMA were analyzed \cite{NZhang,MAI,KH}. Modeling and analysis of conventional uplink NOMA were further evaluated by exploiting various aspects such as user pairing theory \cite{MAS}, power allocation \cite{ZY} and energy harvesting designs \cite{PDD,jiaye}. In terms of GF NOMA networks, code-domain multiplexing is universally considered on multi-user detection designs \cite{BW,YD}, while the research contributions on power domain GF NOMA designs are still in their infancy.

As a powerful mathematical tool to capture the spatial randomness of wireless networks, stochastic geometry has been widely utilized for analyzing the performance of various networks \cite{martin}. For clarify, the stochastic models and distance distributions are evaluated such as homogeneous Poisson point process (HPPP) and Poisson cluster processes (PCP) for cellular networks \cite{JG,HE,Y.JSAC}. With the aid of stochastic geometry methods, some initial NOMA contributions have been investigated \cite{RA,wenqiang2,COMP,UAV1PCP,tianwei,wenqiang,Y.JSAC2}. More particularly, a massive GF NOMA network \cite{RA} and a cache-enabled heterogeneous network \cite{wenqiang2} were recognized as finite uniformly random networks, thereby were investigated by HPPP. For scenarios with nodes in randomly distributed clusters, PCP is universally invoked to model the spatial distributions of clustered nodes, such as coordinated multi-point transmission (CoMP) systems \cite{COMP}, UAV networks \cite{UAV1PCP,tianwei} and clustered millimeter-wave networks \cite{wenqiang}. Moreover, locations of users were arranged into discs and rings in \cite{Y.JSAC2} to simplify the spatial distributions.

Aiming at enhancing the spectral efficiency, related works considered the semi-grant-free NOMA networks \cite{st32,GFGB}, where the GF users are admitted to share the same spectrum resources of the GB users. Hence, both low latency communication and high spectral efficiency are achieved with reliable user experience. More specifically, a hybrid (orthogonal/non-orthogonal) pilot design has set on orthogonal frequency division multiplexing (OFDM) with code domain NOMA in \cite{GFGB}, whereas power domain NOMA has not been investigated. Authors of \cite{st32} have evaluated the outage performance of power domain NOMA assisted networks, whereas the spatial effect of semi-GF NOMA systems has not been investigated. Additionally, two contention control protocols have been proposed with fixed channel gain thresholds. Note that the thresholds were designed for the GF users to access into the channels occupied by the GB users. Nevertheless, a significant requirement is \textit{how to set the value of the channel quality thresholds in stochastic geometry models} as inappropriate thresholds will result in considerate degradations on user experience.

\subsection{Contribution}

Motivated by the aforementioned challenges, we investigate uplink semi-GF NOMA networks where the GF and GB users are combined as one orthogonal pairs, which are employed into the same resource blocks. Since the distances for the GB and GF users are not pre-determined, there are two potential scenarios that: 1) the GF users are located as near users while the GB users as cell-edge users, denoted as \textit{Scenario I} and 2) the GB users are situated in the center areas while the GF users are determined as far users, denoted as \textit{Scenario II}. As pass loss has more stable and dominant effects than instantaneous small-scale fading \cite{SIC_large}, near users always have the best channel gains with the first SIC order than far ones. Based on the mentioned scenarios, the primary contributions are summarized as:

\begin{itemize}
  \item We propose a novel \textit{dynamic protocol} to determine whether the GF users can join into the occupied channels by the GB users. Compared with open-loop protocol, more accurate channel quality thresholds are provided by dynamic protocol, thereby the unexpected interference from the GF users is reduced. We invoke stochastic geometry to exploit the spatial effects of the considered semi-GF NOMA networks. We derive new statistics for combined channels gains of investigated networks.
  \item \textit{For Scenario I}: we derive analytical expressions of outage probability (OP) for the GF and GB users under two protocols. Furthermore, we derive diversity orders for the GF and GB users by carrying out asymptotic analysis in this semi-GF NOMA network. Analytical results reveal that two protocols, i.e., open-loop protocol and dynamic protocol, have the same diversity gains.
  \item \textit{For Scenario II}: we derive the analytical and asymptotic expressions of OP when the SIC orders turn out the contrary compared with Scenario I. We additionally evaluate the diversity gains for the GF and GB users. Our results illustrate that the diversity orders are determined by the SIC orders that: 1) the value equals to one for near users and 2) zero for far users.
  \item Simulation results demonstrate the observation for two scenarios that dynamic protocol outperforms open-loop protocol since less interference from the GF users are involved by dynamic protocol.
\end{itemize}

\subsection{Organization}
The rest sections of this paper are organized as follows. In Section II, the system model for the semi-grant-free NOMA network is introduced. In Section III as Scenario I and in section IV as Scenario II, the outage performance of users is analyzed with diversity orders as valuable insights under open-loop protocol and dynamic protocol respectively. Numerical results are indicated in Section V, followed by the conclusions in Section VI.

\section{Network model}
\begin{figure*}[t]
\centering
\includegraphics[width= 7in]{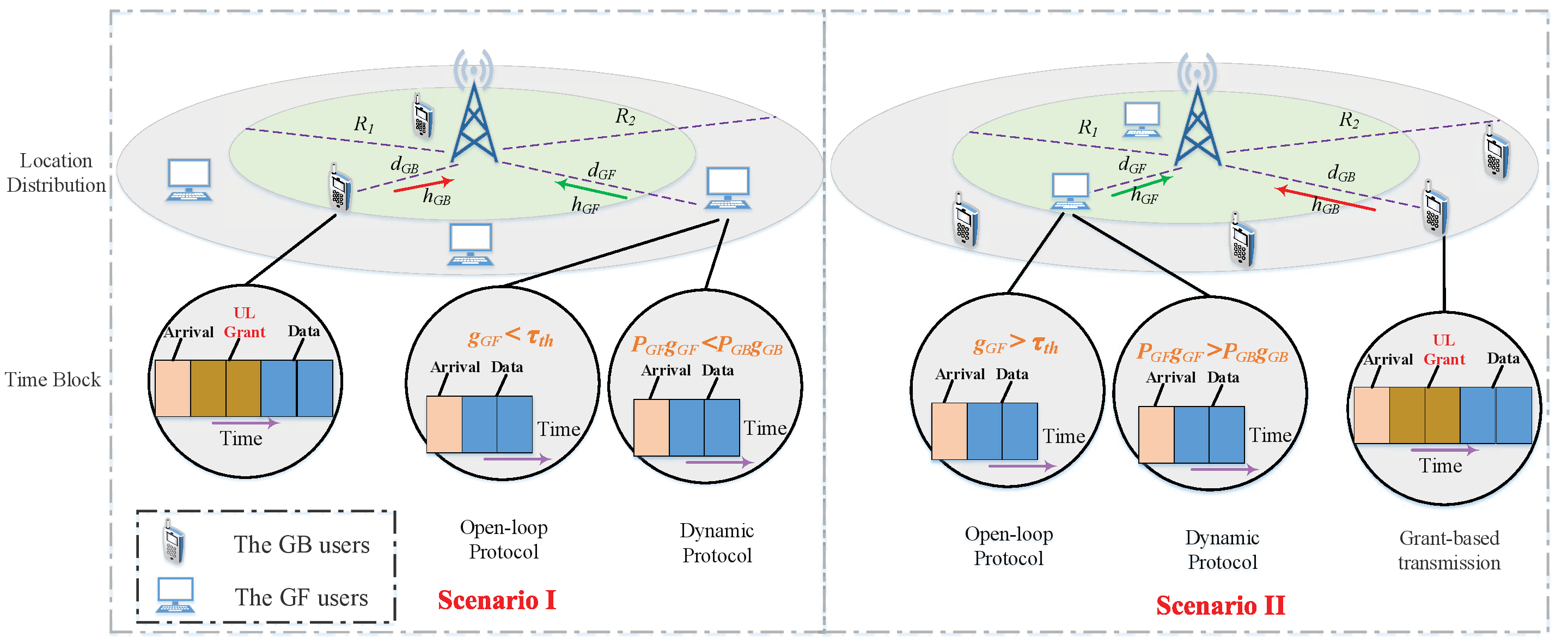}
\caption{An illustration of uplink NOMA networks for conventional grant-base transmission, open-loop semi-grant-free protocol and dynamic semi-grant-free protocol.}
\label{fig_1}
\hrulefill
\end{figure*}
\begin{figure*}[t]
\centering
\includegraphics[width= 7in]{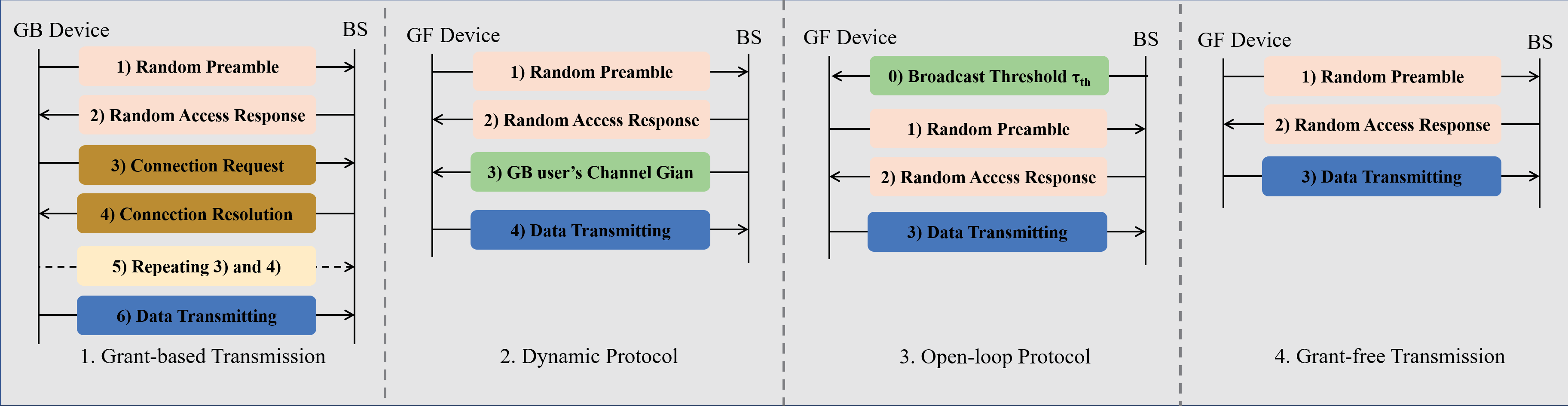}
\caption{Illustrations of handshakes for conventional grant-based transmission, conventional grant-free transmission, open-loop semi-grant-free protocol and dynamic semi-grant-free protocol.}
\label{handshakes}
\hrulefill
\end{figure*}

\subsection{Network Description}

On uplink transmission scenarios as Fig. \ref{fig_1}, we consider semi-GF NOMA networks with $N_{GF}$ GF users and $N_{GB}$ GB users. When GF users follow a certain protocol, i.e., open-loop protocol and dynamic protocol, the GF users can be admitted into the channels occupied by the GB users. We assume $N_{GF}=N_{GB}=M$, thus $M$ pairs of users are randomly combined by one of the GF users and one of the GB users, which means each pair can be allocated into the same orthogonal channel resources in a time block without interferences from other pairs. For simplicity, we draw our attention to the performance of a typical pair of users in this treatise.

Considered that the BS is fixed at the center of the disc, two types of spatial distributions of users are investigated as Fig. \ref{fig_1} that: 1) in \emph{Scenario I}, the GB users $\{GB_j\}$ as the near users are randomly located into the disc with the radium as $R_1$ m and the GF users $\{GF_i\}$ as the far users are deployed within the ring with radius $R_1$ and $R_2$ (assuming $R_2>R_1$), 2) while in \emph{Scenario II}, the GF users are in the disc and the GB users are in the ring. We model the location distributions of the GF and GB users as two HPPPs ${\Phi _{GF}}$ and ${\Phi _{GB}}$ with densities ${\lambda _{{\Phi _{GF}}}}$ and ${\lambda _{{\Phi _{GB}}}}$. Thus, the number of the GF and GB users obey the Poisson distribution, which are expressed as $\Pr \left( {{N_G} = k} \right) = \left( {{{\mu _G^k} \mathord{\left/
 {\vphantom {{\mu _G^k} {k!}}} \right.
 \kern-\nulldelimiterspace} {k!}}} \right)\exp \left( { - \mu _G} \right)$, where $G \in \{ {GF,GB} \}$, $\mu _G$ denotes the mean measures for the GF and GB users, i.e., $\mu _G=\pi R_1^2{\lambda _{{\Phi _G}}}$ for the near users and $\mu _G=\pi \left( {R_2^2 - R_1^2} \right){\lambda _{{\Phi _G}}}$ for the far users. We assume that all users obey independently identically distributions (i.i.d) and are uniformly distributed in the disc and ring. Additionally, we define the distances from the BS to the GF users and the GB users as $d_{GF,i}$ and $d_{GB,j}$, respectively. Hence, the probability density function (PDF) of distances can be derived as ${{{f_{d_{G,k}^{near}}}\left( x \right){\rm{ = 2}}x} \mathord{\left/
 {\vphantom {{{f_{d_{G,k}^{near}}}\left( x \right){\rm{ = 2}}x} {R_1^2}}} \right.
 \kern-\nulldelimiterspace} {R_1^2}}$ and ${{{f_{d_{G,k}^{far}}}\left( x \right){\rm{ = 2}}x} \mathord{\left/
 {\vphantom {{{f_{d_{G,k}^{far}}}\left( x \right){\rm{ = 2}}x} {(R_2^2 - R_1^2)}}} \right.
 \kern-\nulldelimiterspace} {(R_2^2 - R_1^2)}}$, where $k \in \{ i,j\}$.

\subsubsection{The conventional GB and GF transmissions}
According to Fig. \ref{fig_1} and Fig. \ref{handshakes}, the GF and GB transmission schemes are indicated by time blocks and handshakes. For the GB transmission, synchronization is achieved by two handshakes, followed by the transmissions of uplink grants. After experienced a contention access period with at least a pair of handshakes, data transmissions begin. Based on the GB transmission scheme, uplink grants can reduce the collision situations, whereas it culminates long latency. For the GF transmissions, data is transmitted after synchronization without any uplink grants, thereby low latency communications can be achieved. The key challenge is how to satisfy the surge of devices in limited spectrum resources, which causes frequent collision situations.
\vspace{-1mm}
\subsubsection{Open Loop Protocol}
Compared to the conventional GF transmission scheme, we aim at fewer collisions and higher spectrum efficiency. Thus, we consider semi-GF NOMA networks where the GB and GF users share the same spectrum resources in NOMA pairs. Based on the design of open-loop protocol \cite{st32,semi25}, a channel quality threshold $\tau_{th}$ is broadcasted before the transmissions begin, followed by a comparison between the channel gains of the GF users and $\tau_{th}$. The SIC decoding orders determines which portion of the GF users are employed into the channels occupied by the GB users. If the GB users have superior channel gains than the GF users, the GB users would have the first SIC orders, thereby only the GF users with lower channel gains than $\tau_{th}$ will access into the GB channels. When the GB users are decoded with last SIC orders, the GF users with higher channel gains than $\tau_{th}$ are selected into NOMA pairs.

\subsubsection{Dynamic Protocol}
We propose a dynamic protocol to define more accurate values of channel quality thresholds in stochastic geometry models compared to open-loop protocol. We define the combined channel gain of the GB users as ${g_{GB,j}} = {\left| {{h_{GB,j}}} \right|^2}{\left( {{d_{GB.j}}} \right)^{ - \alpha }}$ and the transmit power of the GB users as $P_{GB}$. Compared to open-loop protocol, the key difference of dynamic protocol is that the BS sends various thresholds, denoted as $P_{GB}g_{GB,j}$, for users with different locations by a handshake instead of a fixed threshold $\tau_{th}$ for all users under the open-loop protocol. Note that the comparison between the channel gain of the GF users and access thresholds under dynamic protocol is more accurate than open-loop protocol. Hence, the dynamic protocol is superior to the open-loop protocol on avoiding unexpected interference from the GF users. Moreover, the selection approaches of the GF users to the GB channels are the same as the open-loop protocol.

\subsection{Signal Model}
We model channels of the GF and GB users as Rayleigh fading channels. Based on spatial distributions in two scenarios, the GB users are decoded firstly in Scenario I while it is decoded at the last stage of SIC orders in Scenario II. Hence, we express the SNR expressions in two scenarios.
\subsubsection{Scenario I}
The GB users are deployed within the disc as near users while the GF users are located in the ring as far users. Thus, the GF users have the first SIC decoding orders. With the fixed access thresholds $\tau_{th}$ for open-loop protocol or flexible thresholds $P_{GB}g_{GB,j}$ for dynamic protocol, the SNR of the GB users can be expressed as:
\begin{align}\label{gamma_GB}
\gamma _{GB,j}^I = \frac{{{P_{GB}}{{\left| {{h_{GB,j}}} \right|}^2}\left( {{d_{GB,j}}} \right)^{ - \alpha }}}{{{P_{GF}}{{\left| {{h_{GF,i}}} \right|}^2}\left( {{d_{GF,i}}} \right)^{ - \alpha } + {\sigma ^2}}},
\end{align}
where $P_{GB}$ and $P_{GF}$ are the transmit powers of GB and GF devices, $h_{GF,i} $ and ${h_{GB,j}}$ are the channel gains for $i^{th}$ GF and and $j^{th}$ GB users respectively, $\sigma^{2}$ means variance of additive white Gaussian noise (AWGN) and $\alpha$ is the path loss exponent.

After the cancellation of SIC, the SNR of the GF users can be written as
\begin{align}\label{gamma_GF}
{\gamma^I _{GF,i}} = \frac{P_{GF}{\left| {{h_{GF,i}}} \right|^2}{\left( {{d_{GF,i}}} \right)^{ - \alpha }}}{\sigma^{2} }.
\end{align}
\subsubsection{Scenario II}
The GF users are located in the disc as the near users with first decoding orders in Scenario II, thereby the SNR expressions of the GF and the GB users can be derived respectively as:
\begin{align}\label{gamma_GB2}
{\gamma ^{II}_{GF,i}} = \frac{{{P_{GF}}{{\left| {{h_{GF,i}}} \right|}^2}{{\left( {{d_{GF,i}}} \right)}^{ - \alpha }}}}{{{P_{GB}}{{\left| {{h_{GB,j}}} \right|}^2}{{\left( {{d_{GB,j}}} \right)}^{ - \alpha }} + {\sigma ^2}}}
\end{align}
and
\begin{align}\label{gamma_GF2}
{\gamma ^{II}_{GB,j}} = \frac{{{P_{GB}}{{\left| {{h_{GB,j}}} \right|}^2}{{\left( {{d_{GB,j}}} \right)}^{ - \alpha }}}}{{{\sigma ^2}}}.
\end{align}

Additionally, we simplify the expressions by transmit SNR of users as ${\rho _{GB}} = {{{P_{GB}}} \mathord{\left/
 {\vphantom {{{P_{GB}}} {{\sigma ^2}}}} \right.
 \kern-\nulldelimiterspace} {{\sigma ^2}}}$ and ${\rho _{GF}} = {{{P_{GF}}} \mathord{\left/
 {\vphantom {{{P_{GF}}} {{\sigma ^2}}}} \right.
 \kern-\nulldelimiterspace} {{\sigma ^2}}}$, the combined channel gains as ${g_{GF,i}} = {\left| {{h_{GF,i}}} \right|^2}{\left( {{d_{GF,i}}} \right)^{ - \alpha }}$ for the $i^{th}$ GF user and ${g_{GB,j}} = {\left| {{h_{GB,j}}} \right|^2}{\left( {{d_{GB.j}}} \right)^{ - \alpha }}$ for the $j^{th}$ GB user used.
\subsection{New Statistics}

We combine the channel gains with small-scale and large-scale fading, denoted as $g_{G,k}$ with $G \in \left\{ {GF,GB} \right\}$ and $k \in \left\{ {i,j} \right\}$. \textbf{Lemma \ref{lemma:h/d}} presents derivations of the PDFs of $g_{G,k}$. \textbf{Corollary \ref{CDF_h/d_1}} and \textbf{Corollary \ref{CDF_h/d_2}} derive two types of expressions for CDFs of $g_{G,k}$.

\begin{lemma}\label{lemma:h/d}
\emph{Conditioned on $G \in \left\{ {GF,GB} \right\}$ and $k \in \left\{ {i,j} \right\}$ to express a general scenario for the $j^{th}$ GB and the $i^{th}$ GF users, the combined channel gain concluding large-scale and small-scale fadings is denoted as $g_{G,k}$, whose PDFs for both near and far users can be derived as}
\begin{align} \label{pdf_h/d_near}
f_{{g_{G,k}}}^{near}\left( x \right) = \frac{{b_{1,1}^G}}{{{x^{{b_3}}}}}\gamma \left( {{b_3},b_{2,1}^Gx} \right)
\end{align}
\emph{and}
\begin{align} \label{pdf_h/d_far}
f_{{g_{G,k}}}^{far}\left( x \right) = \frac{{b_{1,2}^G}}{{{x^{{b_3}}}}}\left[ {\gamma \left( {{b_3},b_{2,2}^Gx} \right) - \gamma \left( {{b_3},b_{2,1}^Gx} \right)} \right],
\end{align}
\emph{where $\gamma \left( { \cdot , \cdot } \right)$ means lower incomplete gamma function, ${\lambda _G}$ is the mean of Rayleigh distribution with $G= GB$ for the GB users and $G=GF$ for the GF users, $b_{1,1}^G = {{2{{\left( {{\lambda _G}} \right)}^{\frac{2}{\alpha }}}} \mathord{\left/
 {\vphantom {{2{{\left( {{\lambda _G}} \right)}^{\frac{2}{\alpha }}}} {\alpha R_1^2}}} \right.
 \kern-\nulldelimiterspace} {\alpha R_1^2}}$, $b_{1,2}^G = {{2{{\left( {{\lambda _G}} \right)}^{\frac{2}{\alpha }}}} \mathord{\left/
 {\vphantom {{2{{\left( {{\lambda _G}} \right)}^{\frac{2}{\alpha }}}} {\left[ {\alpha \left( {R_2^2 - R_1^2} \right)} \right]}}} \right.
 \kern-\nulldelimiterspace} {\left[ {\alpha \left( {R_2^2 - R_1^2} \right)} \right]}}$, $b_{2,1}^G = {{R_1^\alpha } \mathord{\left/
 {\vphantom {{R_1^\alpha } {{\lambda _G}}}} \right.
 \kern-\nulldelimiterspace} {{\lambda _G}}}$, $b_{2,2}^G = {{R_2^\alpha } \mathord{\left/
 {\vphantom {{R_2^\alpha } {{\lambda _G}}}} \right.
 \kern-\nulldelimiterspace} {{\lambda _G}}}$ and $b_3=\frac{2}{\alpha }+1$.}
\begin{IEEEproof}
Near users and far users are deployed into the disc or the ring. Based on the PDFs of ${ d_{G,k} }$, the PDFs of ${ d_{G,k} ^\alpha }$ can be derived as ${f_{{{\left( {d_{GB,i}^{near}} \right)}^\alpha }}}\left( x \right) = {{2{x^{\frac{2}{\alpha } - 1}}} \mathord{\left/
 {\vphantom {{2{x^{\frac{2}{\alpha } - 1}}} {\left( {\alpha R_1^2} \right)}}} \right.
 \kern-\nulldelimiterspace} {\left( {\alpha R_1^2} \right)}}$ and ${f_{{{\left( {d_{GB,i}^{far}} \right)}^\alpha }}}\left( x \right) = {{2{x^{\frac{2}{\alpha } - 1}}} \mathord{\left/
 {\vphantom {{2{x^{\frac{2}{\alpha } - 1}}} {\left[ {\alpha \left( {R_2^2 - R_1^2} \right)} \right]}}} \right.
 \kern-\nulldelimiterspace} {\left[ {\alpha \left( {R_2^2 - R_1^2} \right)} \right]}}$. Under Rayleigh fading channels, the PDFs of $g_{G,k}$ can be derived as
\begin{align}\label{repdf_h/d}
f_{{g_{G,k}}}^{near}\left( x \right) = \int_0^{R_1^\alpha } {y{f_{{{\left| {{h_{G,k}}} \right|}^2}}}} \left( {xy} \right){f_{{{\left( {d_{G,k}^{near}} \right)}^\alpha }}}\left( y \right)dy
\end{align}
and
\begin{align}\label{repdf_h/d}
f_{{g_{G,k}}}^{far}\left( x \right) = \int_{R_1^\alpha }^{R_2^\alpha } {y{f_{{{\left| {{h_{G,k}}} \right|}^2}}}} \left( {xy} \right){f_{{{\left( {d_{G,k}^{far}} \right)}^\alpha }}}\left( y \right)dy,
\end{align}
which can be derived by substituting the PDFs of Exponential distribution and ${ d_{G,k} ^\alpha }$ as \eqref{pdf_h/d_near} and \eqref{pdf_h/d_far}.
\end{IEEEproof}
\end{lemma}

\begin{corollary}\label{CDF_h/d_1}
\emph{Based on the PDF of $g_{G,k}$ in \textbf{Lemma \ref{lemma:h/d}}, expressions of the the CDFs of $g_{G,k}$ can be derived by utilizing hypergeometric functions as }
\begin{align}\label{CDF_near1}
F_{{g_{G,k}}}^{near}(x) = \frac{{b_{1,1}^G{{\left( {b_{2,1}^G} \right)}^{{b_3}}}}}{{{b_3}}}x{}_2{F_2}\left( {{b_3},1;{b_3} + 1,2; - b_{2,1}^Gx} \right)
\end{align}
\emph{and}
\begin{align}\label{CDF_far1}
F_{{g_{G,k}}}^{far}(x) = \frac{{b_{1,2}^G{{\left( {b_{2,2}^G} \right)}^{{b_3}}}}}{{{b_3}}}x{}_2{F_2}\left( {{b_3},1;{b_3} + 1,2; - b_{2,2}^Gx} \right)\notag\\
 - \frac{{b_{1,2}^G{{\left( {b_{2,1}^G} \right)}^{{b_3}}}}}{{{b_3}}}x{}_2{F_2}\left( {{b_3},1;{b_3} + 1,2; - b_{2,1}^Gx} \right),
\end{align}
\emph{where ${}_p{F_q}\left( { \cdot  } \right)$ is the hypergeometric function.}
\begin{IEEEproof}
In terms of the expressions of Eq.[2.10.2.2] in \cite{Prudnikov2}, the CDFs can be derived via the PDFs in \textbf{Lemma \ref{lemma:h/d}}.
\end{IEEEproof}
\end{corollary}

\begin{corollary}\label{CDF_h/d_2}
\emph{Invoked by the lower incomplete gamma functions and the CDF of Exponential distribution, the CDF expressions of $g_{G,k}$ can be equivalently derived as}
\begin{align}\label{CDF_near2}
F_{{g_{G,k}}}^{near}(x) = 1 - \frac{2}{{\alpha R_1^2}}{\left( {\frac{{{\lambda _G}}}{x}} \right)^{{b_3} - 1}}\gamma \left( {{b_3} - 1,b_{2,1}^Gx} \right)
\end{align}
\emph{and}
\begin{align}\label{CDF_far2}
F_{{g_{G,k}}}^{far}(x)&=1 - \frac{2}{{\alpha \left( {R_2^2 - R_1^2} \right)}}{\left( {\frac{{{\lambda _G}}}{x}} \right)^{{b_3} - 1}}\notag\\
 &\hspace*{0.3cm} \times \left[ {\gamma \left( {{b_3} - 1,b_{2,2}^Gx} \right) - \gamma \left( {{b_3} - 1,b_{2,1}^Gx} \right)} \right].
\end{align}
\begin{IEEEproof}
The CDFs of $g_{G,k}$ can be expressed as $ F_{{g_{G,k}}}^{near}(x) = \Pr \left\{ {{{\left| {{h_{G,k}}} \right|}^2} < {{\left( {d_{G,k}^{near}} \right)}^\alpha }x} \right\} $ for the near users and $F_{{g_{G,k}}}^{far}(x){\rm{ = }}\Pr \left\{ {{{\left| {{h_{G,k}}} \right|}^2} < {{\left( {d_{G,k}^{far}} \right)}^\alpha }x} \right\}$ for the far users. Thus, the CDF expressions can be derived as \eqref{CDF_near2} and \eqref{CDF_far2}.
\end{IEEEproof}
\end{corollary}

\section{Outage Performance in Scenario I}

In Scenario I, we assume the GB users as the near users and the GF users as the far users to define the SIC orders. Considered that a pair of NOMA users with a GF user and a GB user are randomly selected, the outage performance of the GF and GB users are analyzed under two semi-GF protocols. We express the exact expressions of OP via \textbf{Theorem \ref{OPGB2}} to \textbf{Theorem \ref{OPGF1}}, followed by several corollaries to derive the closed-form expressions.

\subsection{Analytical OP under Dynamic Protocol in Scenario I}
The channel capacities of the GB and GF users are given by ${C_{G{B_j}}} = {\log _2}\left( {1 + \gamma _{GB,j}^I} \right) = {\log _2}\left( {1 + \frac{{{\rho _{GB}}{g_{GB,j}}}}{{{\rho _{GF}}{g_{GF,i}} + 1}}} \right)$ and ${C_{G{F_j}}} = {\log _2}\left( {1 + \gamma _{GF,i}^I} \right) = {\log _2}\left( {1 + {\rho _{GF}}{g_{GF,i}}} \right)$. Conditioned on $P_{GF}g_{GF,i}<P_{GB}g_{GB,j}$ in Scenario I, the OP expressions of the GF and the GB users are expressed as
\begin{align}\label{P_out_s2_GB}
P_{out,{p_2}}^{GB,I} = \Pr \left\{ {\frac{{{\rho _{GB}}{g_{GB,j}}}}{{{\rho _{GF}}{g_{GF,i}} + 1}} < \gamma _{th}^{GB},{g_{GF,i}} < \frac{{{P_{GB}}}}{{{P_{GF}}}}{g_{GB,j}}} \right\}
\end{align}
and
\begin{align}\label{P_out_s2_GF}
P_{out,{p_2}}^{GF,I}& = P_{out,{p_2}}^{GB,I} + \Pr \left\{ {\frac{{{\rho _{GB}}{g_{GB,j}}}}{{{\rho _{GF}}{g_{GF,i}} + 1}} > \gamma _{th}^{GB},} \right.\notag\\
&\hspace*{0.3cm} \left. {{g_{GF,i}} < \frac{{\gamma _{th}^{GF}}}{{{\rho _{GF}}}},{g_{GF,i}} < \frac{{{P_{GB}}{g_{GB,j}}}}{{{P_{GF}}}}} \right\},
\end{align}
where the second item of the probabilities in \eqref{P_out_s2_GF} is denoted as $Q_1$ to simplify the expressions.

\begin{theorem}\label{OPGB2}
\emph{Conditioned on $P_{GF}{{g_{GF,i}} < P_{GB}{g_{GB,j}}}$ under dynamic protocol in Scenario I, the OP of the GB users varies into two situations with various derivations: \textit{a)} the outage threshold of the GB users is higher than one, denoted as $\gamma _{th}^{GB} > 1$ and \textit{b)} the other situation is that the system has a low outage threshold of the GB user, denoted as $\gamma _{th}^{GB} \le 1$. Based on two situations, the OP of the GB users can be derived as}
\begin{align}
&P_{out,{p_2}}^{GB,I} = \int_0^\infty  {F_{{g_{GB,j}}}^{near}\left( {\frac{{\gamma _{th}^{GB}{\rho _{GF}}x + \gamma _{th}^{GB}}}{{{\rho _{GB}}}}} \right)} f_{{g_{GF,i}}}^{far}\left( x \right)dx\notag\\
&\hspace*{0.3cm}\begin{array}{*{20}{c}}
{ - \int_0^\infty  {F_{{g_{GB,j}}}^{near}\left( {\frac{{{\rho_{GF}}}}{{{\rho_{GB}}}}x} \right)} f_{{g_{GF,i}}}^{far}\left( x \right)dx,}&{\left( \gamma _{th}^{GB} > 1 \right)}
\end{array}
\end{align}
\emph{and}
\begin{align}
&P_{out,{p_2}}^{GB,I} = \int_0^{{\sigma _1}} {F_{{g_{GB,j}}}^{near}\left( {\frac{{\gamma _{th}^{GB}{\rho _{GF}}x + \gamma _{th}^{GB}}}{{{\rho _{GB}}}}} \right)} f_{{g_{GF,i}}}^{far}\left( x \right)dx\notag\\
&\hspace*{0.3cm}\begin{array}{*{20}{c}}
{ - \int_0^{{\sigma _1}} {F_{{g_{GB,j}}}^{near}\left( {\frac{{{\rho_{GF}}}}{{{\rho_{GB}}}}x} \right)} f_{{g_{GF,i}}}^{far}\left( x \right)dx,}&{\left( \gamma _{th}^{GB} \le 1 \right)},
\end{array}
\end{align}
\emph{where ${\sigma _1} = {\frac{{\gamma _{th}^{GB}}}{{{\rho_{GF}}\left( {1 - \gamma _{th}^{GB}} \right)}}}$. \textbf{Corollary \ref{AOPGB21}} and \textbf{Corollary \ref{AOPGB23}} can express the closed-form expressions in two situations.}
\end{theorem}

\begin{corollary}\label{AOPGB21}
\emph{Assume that the GF users experience satisfying channel conditions with high transmit SNR $\rho _{GF} \gg 1$. Conditioned on $\gamma _{th}^{GB} > 1$, the exact closed-from expressions of OP for the GB users can be derived as}
\begin{align}
&P_{out,{p_2}}^{GB,I} = {C_1}\left[ {U\left( {\frac{{b_{2,1}^{GB}{\rho _{GF}}}}{{{\rho _{GB}}}},b_{2,2}^{GF}} \right) - U\left( {\frac{{b_{2,1}^{GB}{\rho _{GF}}}}{{{\rho _{GB}}}},b_{2,1}^{GF}} \right)} \right]\notag\\
&\hspace*{0.3cm}  - {C_1}\sum\limits_{n = 0}^\infty  {\sum\limits_{t = 0}^n {\dbinom{n}{r}} {{\left( {\frac{{ - \gamma _{th}^{GB}{\rho _{GF}}}}{{{\rho _{GB}}}}} \right)}^n}} \frac{{{{\left( {b_{2,1}^{GB}} \right)}^{\frac{2}{\alpha } + n}}}}{{n!\left( {\frac{2}{\alpha } + n} \right)}}\notag\\
&\hspace*{0.3cm}  \times \frac{{\Gamma \left( {n - t - \frac{2}{\alpha }} \right)\left[ {{{\left( {b_{2,1}^{GF}} \right)}^{\frac{2}{\alpha } + t - n}} - {{\left( {b_{2,2}^{GF}} \right)}^{\frac{2}{\alpha } + t - n}}} \right]}}{{\rho _{GF}^t\left( {n - t - \frac{2}{\alpha }} \right)}},
\end{align}
\emph{where ${C_1} = {{2\lambda _{GB}^{\frac{{\rm{2}}}{\alpha }}b_{1,2}^{GF}} \mathord{\left/
 {\vphantom {{2\lambda _{GB}^{\frac{{\rm{2}}}{\alpha }}b_{1,2}^{GF}} {\left( {\alpha R_1^2} \right)}}} \right.
 \kern-\nulldelimiterspace} {\left( {\alpha R_1^2} \right)}}$, $\dbinom{n}{r} = {{n!} \mathord{\left/
 {\vphantom {{n!} {\left[ {t!\left( {n - t} \right)!} \right]}}} \right.
 \kern-\nulldelimiterspace} {\left[ {t!\left( {n - t} \right)!} \right]}}$, $U\left( {a,t} \right) ={}_3{F_2}\left( {1,{b_3},2 - {b_3};{b_3} + 1,3 - {b_3};\frac{{ - t}}{a}} \right){\theta _1} - {\theta _2}$, ${\theta _1} = \frac{{ - {t^{{b_3}}}\Gamma \left( 1 \right)}}{{{b_3}\left( {2 - {b_3}} \right){a^{2 - {b_3}}}}}$, ${\theta _2} = \frac{{\Gamma \left( {2 - {b_3}} \right)\Gamma \left( {{b_3} - 1} \right)}}{{2\left( {1 - {b_3}} \right){t^{2\left( {1 - {b_3}} \right)}}}}$ and $\Gamma(\cdot)$ is gamma function. }
\begin{IEEEproof}
Substituting \eqref{pdf_h/d_far} and \eqref{CDF_near2} into the OP expressions, $P_{out,{p_2}}^{GB,I}$ can be rewritten as \eqref{P_GB_2}, which defined $I_1$ and $I_2$.
\begin{figure*}[t]
\begin{align} \label{P_GB_2}
&P_{out,{p_2}}^{GB,I} = \underbrace {\int_0^\infty  {\frac{2}{{\alpha R_1^2}}{{\left( {\frac{{{\lambda _{GB}}}}{x}} \right)}^{{b_3} - 1}}\gamma \left( {{b_3} - 1,b_{2,1}^{GB}x} \right)} \frac{{b_{1,2}^{GF}}}{{{x^{{b_3}}}}}\left[ {\gamma \left( {{b_3},b_{2,2}^{GF}x} \right) - \gamma \left( {{b_3},b_{2,1}^{GF}x} \right)} \right]dx}_{{I_1}}\notag\\
&\hspace*{0.3cm} - \underbrace {\int_0^\infty  {\frac{2}{{\alpha R_1^2}}} {{\left( {\frac{{{\rho _{GB}}{\lambda _{GB}}}}{{\gamma _{th}^{GB}{\rho _{GF}}x + \gamma _{th}^{GB}}}} \right)}^{{b_3} - 1}}\gamma \left[ {{b_3} - 1,\frac{{\left( {{\rho _{GF}}x + 1} \right)\gamma _{th}^{GB}b_{2,1}^{GB}}}{{{\rho _{GB}}}}} \right]\frac{{b_{1,2}^{GF}}}{{{x^{{b_3}}}}}\left[ {\gamma \left( {{b_3},b_{2,2}^{GF}x} \right) - \gamma \left( {{b_3},b_{2,1}^{GF}x} \right)} \right]dx}_{{I_{\rm{2}}}}.
\end{align}
\hrulefill
\end{figure*}

Based on Eq.[2.10.6.2] in \cite{Prudnikov2} to derive $I_1$, the expressions can be simplified as
\begin{align} \label{I1}
{I_1} &= {C_1}\int_0^\infty  {{x^{2\left( {1 - {b_3}} \right) - 1}}\gamma \left( {{b_3} - 1,\frac{{b_{2,1}^{GB}{\rho _{GF}}}}{{{\rho _{GB}}}}x} \right)} \notag\\
 &\hspace*{0.3cm}\times \left[ {\gamma \left( {{b_3},b_{2,2}^{GF}x} \right) - \gamma \left( {{b_3},b_{2,1}^{GF}x} \right)} \right]dx.
\end{align}

Based on the expansions of lower incomplete gamma functions as \eqref{gamma_asy}, binomial expansions and Eq. [2.10.2.1] in \cite{Prudnikov2}, we derive the closed-form expressions of $I_2$ as
\begin{align} \label{I2}
&{I_2} = {C_1}\sum\limits_{n = 0}^\infty  {\sum\limits_{t = 0}^n {\dbinom{n}{r}} {{\left( {\frac{{ - \gamma _{th}^{GB}{\rho _{GF}}}}{{{\rho _{GB}}}}} \right)}^n}} \frac{{{{\left( {b_{2,1}^{GB}} \right)}^{\frac{2}{\alpha } + n}}}}{{n!\left( {\frac{2}{\alpha } + n} \right)}}\notag\\
&\hspace*{0.3cm}  \times \frac{{\Gamma \left( {n - t - \frac{2}{\alpha }} \right)\left[ {{{\left( {b_{2,1}^{GF}} \right)}^{\frac{2}{\alpha } + t - n}} - {{\left( {b_{2,2}^{GF}} \right)}^{\frac{2}{\alpha } + t - n}}} \right]}}{{\rho _{GF}^t\left( {n - t - \frac{2}{\alpha }} \right)}}.
\end{align}

Substituting \eqref{I1} and \eqref{I2} into \eqref{P_GB_2}, the corollary is proved.
\end{IEEEproof}
\end{corollary}

\begin{corollary}\label{AOPGB23}
\emph{Conditioned on ${\omega _s} = \frac{\pi }{N}$ and ${x_s} = \cos \left( {\frac{{2s - 1}}{{2S}}\pi } \right)$, Chebyshev-gauss quadrature as a numerical analytical method with limited upper and lower limits is approximated as $\int_a^b {f\left( x \right)} dx = \sum\limits_{s = 1}^S {\frac{{(b - a){\omega _s}}}{{2{{\left[ {1 - t_s^2\left( {{x_s},a,b} \right)} \right]}^{ - \frac{{\rm{1}}}{{\rm{2}}}}}}}f\left[ {{t_s}\left( {{x_s},a,b} \right)} \right]}  $, where ${t_s}\left( {{x_s},a,b} \right) = \left( {{x_s} + 1} \right)\frac{{b - a}}{2} + a$. When $\gamma _{th}^{GB} \le 1$ as the second situation, the OP expressions of the GB users can be derived as}
\begin{align}
&P_{out,{p_2}}^{GB,I} = \underbrace {\int_0^{{\sigma _{\rm{1}}}} {F_{{g_{GB,j}}}^{near}\left( {\frac{{\gamma _{th}^{GB}{\rho _{GF}}x + \gamma _{th}^{GB}}}{{{\rho _{GB}}}}} \right)} f_{{g_{GF,i}}}^{far}\left( x \right)dx}_{{I_3}}\notag\\
&\hspace*{0.3cm}  - \underbrace {\int_0^{{\sigma _{\rm{1}}}} {F_{{g_{GB,j}}}^{near}\left( {\frac{{{\rho_{GF}}}}{{{\rho_{GB}}}}}x \right)} f_{{g_{GF,i}}}^{far}\left( x \right)dx}_{{I_4}},
\end{align}
\emph{where closed-from expressions of $I_3$ and $I_4$ can be derived as }
\begin{align}
&{I_3} =F_{{g_{GF,i}}}^{far}\left( {{\sigma _{\rm{1}}}} \right) - \sum\limits_{s = 1}^S {{}\gamma \left[ {\frac{2}{\alpha },\frac{{\gamma _{th}^{GB}b_{2,1}^{GB}\left( {{\rho _{GF}}{\iota _{s,1}} + 1} \right)}}{{{\rho _{GB}}}}} \right]}  \notag\\
 &\hspace*{0.3cm} \times {\Lambda _1}\left( {{\sigma _{\rm{1}}},{\iota _{s,1}}} \right)\left[ {\gamma \left( {{b_3},b_{2,2}^{GF}{\iota _{s,1}}} \right) - \gamma \left( {{b_3},b_{2,1}^{GF}{\iota _{s,1}}} \right)} \right]
\end{align}
\emph{and}
\begin{align}
{I_4} &= F_{{g_{GF,i}}}^{far}\left( {{\sigma _{\rm{1}}}} \right) - \sum\limits_{s = 1}^S {{{\Lambda _2}\left( {{\sigma _{\rm{1}}},{\iota _{s,1}}} \right)}\gamma \left( {{b_3} - 1,b_{2,1}^G{\iota _{s,1}}} \right)} \notag\\
 &\hspace*{0.3cm}\times \left[ {\gamma \left( {{b_3},b_{2,2}^{GF}{\iota _{s,1}}} \right) - \gamma \left( {{b_3},b_{2,1}^{GF}{\iota _{s,1}}} \right)} \right],
\end{align}
\emph{where ${\Lambda _2}\left( a,x \right) = \frac{{{C_1}}}{2}{a}{\omega _s}{x^{1 - 2{b_3}}}{\left( {1 - {x^2}} \right)^{\frac{{\rm{1}}}{{\rm{2}}}}}\rho _{GB}^{{b_3} - 1}\rho _{GF}^{1 - {b_3}}$, ${\Lambda _1}\left( {a,x} \right) = \frac{{{C_1}}}{2}a{\omega _s}{x^{ - {b_3}}}{\left( {1 - {x^2}} \right)^{\frac{{\rm{1}}}{{\rm{2}}}}}\rho _{GB}^{{b_3} - 1}{\left[ {\gamma _{th}^{GB}\left( {{\rho _{GF}}x + 1} \right)} \right]^{\frac{-2}{\alpha }}}$ and ${\iota _{s,1}} = {t_s}\left( {{x_s},0,{\sigma _{\rm{1}}}} \right)$.}
\begin{IEEEproof}
Based on \eqref{pdf_h/d_far}, \eqref{CDF_near1}, \eqref{CDF_near2} and Chabyshev-gauss quadrature, we can obtain the closed-form expressions.
\end{IEEEproof}
\end{corollary}

\begin{theorem}\label{OPGF2}
\emph{Note that $P_{out,{p_2}}^{GB,I}$ is given as \textbf{Theorem \ref{OPGB2}} and $Q_1$ is given in \eqref{P_out_s2_GF}. The integration $Q_1$ can be derived by different derivations under two situations: \textit{a)} $\gamma _{th}^{GB} > {{\gamma _{th}^{GF}} \mathord{\left/
 {\vphantom {{\gamma _{th}^{GF}} {\left( {1 + \gamma _{th}^{GF}} \right)}}} \right.
 \kern-\nulldelimiterspace} {\left( {1 + \gamma _{th}^{GF}} \right)}}$ and \textit{b)} $\gamma _{th}^{GB} \le {{\gamma _{th}^{GF}} \mathord{\left/
 {\vphantom {{\gamma _{th}^{GF}} {\left( {1 + \gamma _{th}^{GF}} \right)}}} \right.
 \kern-\nulldelimiterspace} {\left( {1 + \gamma _{th}^{GF}} \right)}}$. Thus, The OP of the GF users under dynamic protocol in Scenario I can be expressed as }
\begin{align}
P_{out,{p_2}}^{GF,I} = {Q_1} + P_{out,{p_2}}^{GB,I},
\end{align}
\emph{where we can achieve the closed-from expressions by substituting $P_{out,{p_2}}^{GB,I}$ in \textbf{Theorem \ref{OPGB2}} and $Q_1$ in the following corollaries as \textbf{Corollary \ref{AOPGF21}}, \textbf{Corollary \ref{AOPGF22}} and \textbf{Corollary \ref{AOPGF23}}.}
\end{theorem}

\begin{corollary}\label{AOPGF21}
\emph{Note that the first situation as $\gamma _{th}^{GB} > {{\gamma _{th}^{GF}} \mathord{\left/
 {\vphantom {{\gamma _{th}^{GF}} {\left( {1 + \gamma _{th}^{GF}} \right)}}} \right.
 \kern-\nulldelimiterspace} {\left( {1 + \gamma _{th}^{GF}} \right)}}$ is considered. For the first case that we use the expansions of lower incomplete gamma functions and binomial expansions, the closed-from expressions of $Q_1$ can be derived as }
\begin{align}
{Q_1} &= {C_1}\sum\limits_{n = 0}^\infty  {\sum\limits_{t = 0}^n { {\dbinom{n}{r}} } {{\left( {\frac{{ - \gamma _{th}^{GB}{\rho _{GF}}}}{{{\rho _{GB}}}}} \right)}^n}} \frac{{{{\left( {b_{2,1}^{GB}} \right)}^{\frac{2}{\alpha } + n}}}}{{n!\left( {\frac{2}{\alpha } + n} \right)\rho _{GF}^t}}\notag\\
 &\hspace*{0.3cm} \times \left[ {M\left( {\sigma _2,q,{b_3},b_{2,2}^{GF}} \right) - M\left( {\sigma _2,q,{b_3},b_{2,1}^{GF}} \right)} \right],
\end{align}
\emph{where $\sigma _2={{\gamma _{th}^{GF}} \mathord{\left/
 {\vphantom {{\gamma _{th}^{GF}} {{\rho _{GF}}}}} \right.
 \kern-\nulldelimiterspace} {{\rho _{GF}}}}$, $q = n - t - \frac{2}{\alpha }$ and $M\left( {t,\alpha ,\beta ,\delta } \right)$ is defined as }
\begin{align}
&M\left( {t,\alpha ,\beta ,\delta } \right) = \int_0^t {{x^{\alpha  - 1}}} \gamma \left( {\beta ,\delta x} \right)dx\notag\\
&  = \frac{{{t^{\alpha  + \beta }}}}{{{\delta ^{ - \beta }}\beta }}B{\left( {1,\alpha  + \beta } \right)_2}{F_2}\left( {\beta ,\alpha  + \beta ;\beta  + 1,\alpha  + \beta  + 1; - t\delta } \right).
\end{align}
\begin{IEEEproof}
Using $\gamma \left( {a,b} \right) = \sum\limits_{n=0}^\infty  {\frac{{{{\left( { - 1} \right)}^n}{b^{a + n}}}}{{n!\left( {a + n} \right)}}} $, binary series expansions and Eq. [2.10.2.2] in \cite{Prudnikov2}, the corollary is proved.
\end{IEEEproof}
\end{corollary}

\begin{corollary}\label{AOPGF22}
\emph{Note that $\gamma _{th}^{GB} >{{\gamma _{th}^{GF}} \mathord{\left/
 {\vphantom {{\gamma _{th}^{GF}} {\left( {1 + \gamma _{th}^{GF}} \right)}}} \right.
 \kern-\nulldelimiterspace} {\left( {1 + \gamma _{th}^{GF}} \right)}}$ is considered. Chebushev-gauss quadrature is invoked to calculate the closed-form expressions of OP for the GF users, thereby the approximated expressions of $Q_1$ can be presented as }
\begin{align}
{{\rm{Q}}_1} &= \sum\limits_{s = 1}^S {{\Lambda _1\left( {{\sigma _2 ,\iota _{s,2}}} \right)}} \gamma \left[ {{b_3} - 1,\frac{{b_{2,1}^{GB}\gamma _{th}^{GB}\left( {{\rho _{GF}}{\iota _{s,2}} + 1} \right)}}{{{\rho _{GB}}}}} \right]\notag\\
&\hspace*{0.3cm}  \times \left[ {\gamma \left( {{b_3},b_{2,2}^{GF}{\iota _{s,2}}} \right) - \gamma \left( {{b_3},b_{2,1}^{GF}{\iota _{s,2}}} \right)} \right],
\end{align}
\emph{where ${\iota _{s,2}} = {t_s}\left( {{x_s},0,{\sigma _2}} \right)$.}
\end{corollary}

\begin{corollary}\label{AOPGF23}
\emph{Conditioned on the second situation, denoted as $\gamma _{th}^{GB} \le {{\gamma _{th}^{GF}} \mathord{\left/
 {\vphantom {{\gamma _{th}^{GF}} {\left( {1 + \gamma _{th}^{GF}} \right)}}} \right.
 \kern-\nulldelimiterspace} {\left( {1 + \gamma _{th}^{GF}} \right)}}$, the closed-from expressions of $Q_1$ can be derived as   }
\begin{align}
{Q_1} & = \sum\limits_{s = 1}^S {\gamma \left( {{b_3} - 1,b_{2,1}^{GB}\frac{{\gamma _{th}^{GB}{\rho _{GF}}{\iota _{s,1}} + \gamma _{th}^{GB}}}{{{\rho _{GB}}}}} \right)}\notag \\
 &\hspace*{0.3cm}\times {\Lambda _1}\left( {{\sigma _{\rm{1}}},{\iota _{s,1}}} \right)\left[ {\gamma \left( {{b_3},b_{2,2}^{GF}{\iota _{s,1}}} \right) - \gamma \left( {{b_3},b_{2,1}^{GF}{\iota _{s,1}}} \right)} \right]\notag\\
 &\hspace*{0.3cm}+ {\Lambda _2}\left[ {({\sigma _2} - {\sigma _1}),{\iota _{s,12}}} \right]\gamma \left( {{b_3} - 1,\frac{{b_{2,1}^{GB}{\rho _{GF}}}}{{{\rho _{GB}}}}{\iota _{s,12}}} \right)\notag\\
&\hspace*{0.3cm} \times \left[ {\gamma \left( {{b_3},b_{2,2}^{GF}{\iota _{s,12}}} \right) - \gamma \left( {{b_3},b_{2,1}^{GF}{\iota _{s,12}}} \right)} \right],
\end{align}
\emph{where ${\iota _{s,12}} = {t_s}\left( {{x_s},{\sigma _{\rm{1}}},{\sigma _2}} \right)$}.
\end{corollary}

\subsection{Analytical OP under Open-loop Protocol in Scenario I}
Recall that the fixed channel quality thresholds are the average channel gains of the GB users, which are broadcasted under open-loop protocol. Thus, the outage probability of GB and GF users can be expressed as
\begin{align}\label{P_out_s1_GB}
P_{out,{p_1}}^{GB,I} = \Pr \left\{ {\frac{{{\rho _{GB}}{g_{GB,j}}}}{{{\rho _{GF}}{g_{GF,i}} + 1}} < \gamma _{th}^{GB},{g_{GF,i}} < {\tau _{th}}} \right\}
\end{align}
and
\begin{align}\label{P_out_s1_GF}
&P_{out,{p_1}}^{GF,I} = P_{out,{p_1}}^{GB,I}\notag\\
&\hspace*{0.3cm}+\underbrace {\Pr \left\{ {\frac{{{\rho _{GB}}{g_{GB,j}}}}{{{\rho _{GF}}{g_{GF,i}} + 1}} > \gamma _{th}^{GB},{g_{GF,i}} < \min \left( {\frac{{\gamma _{th}^{GF}}}{{{\rho _{GF}}}},{\tau _{th}}} \right)} \right\}}_{{Q_2}}.
\end{align}

\begin{theorem}\label{OPGB1}
\emph{Recall that under open-loop protocol in Scenario I, the GF users with lower channel gain than $\tau_{th}$ are employed into NOMA pairs, denoted as $g_{GF}<\tau_{th}$. With the aforementioned requirement, the outage probability of the GB users can be derived as }
\begin{align}\label{T1}
P_{out,{p_1}}^{GB,I} = \int_0^{{\tau _{th}}} {F_{{g_{GB,j}}}^{near}} \left( {\frac{{\gamma _{th}^{GB}{\rho _{GF}}x + \gamma _{th}^{GB}}}{{{\rho _{GB}}}}} \right)f_{{g_{GF,i}}}^{far}\left( x \right)dx ,
\end{align}
\emph{where the closed-form expressions are derived by \textbf{Corollary \ref{COPGB1}} and \textbf{Corollary \ref{COPGB2}}.}
\end{theorem}

\begin{corollary}\label{COPGB1}
\emph{Based on two types of the CDF expressions of $g_{G,k}$, two types of closed-form expressions by Chebyshev-gauss quadrature can be derived as}
\begin{align}
&P_{out,{p_1}}^{GB,I}{\rm{ = }}\sum\limits_{s = 1}^S {\frac{{{\tau _{th}}b_{1,1}^{GB}b_{1,2}^{GF}{{\left( {b_{2,1}^{GB}} \right)}^{{b_3}}}{\omega _s}}}{{2{{\left( {1 - t_s^2\left( {{x_s},0,{\tau _{th}}} \right)} \right)}^{ - \frac{{\rm{1}}}{{\rm{2}}}}}{b_3}{x^{{b_3} - 1}}}}} \notag\\
 &\times _2{F_2}\left[ {{b_3},1;{b_3} + 1,2;b_{2,1}^{GB}\gamma _{th}^{GB}\left( {\frac{{{\rho _{GF}}{t_s}\left( {{x_s},0,{\tau _{th}}} \right) + {\rm{1}}}}{{ - {\rho _{GB}}}}} \right)} \right]\notag\\
 &\times \left\{ {\gamma \left[ {{b_3},b_{2,2}^{GF}{t_s}\left( {{x_s},0,{\tau _{th}}} \right)} \right] - \gamma \left[ {{b_3},b_{2,1}^{GF}{t_s}\left( {{x_s},0,{\tau _{th}}} \right)} \right]} \right\}
\end{align}
\emph{and}
\begin{align}
&P_{out,{p_1}}^{GB,I} = \sum\limits_{s = 1}^S {\frac{{{\tau _{th}}{\omega _s}{{\left[ {1 - t_s^2\left( {{x_s},0,{\tau _{th}}} \right)} \right]}^{\frac{{\rm{1}}}{{\rm{2}}}}}b_{1,2}^{GF}}}{{2{{\left[ {{t_s}\left( {{x_s},0,{\tau _{th}}} \right)} \right]}^{{b_3}}}}}}\notag \\
&\hspace*{0.3cm}  \times \left\{ {\gamma \left[ {{b_3},b_{2,2}^{GF}{t_s}\left( {{x_s},0,{\tau _{th}}} \right)} \right] - \gamma \left[ {{b_3},b_{2,1}^{GF}{t_s}\left( {{x_s},0,{\tau _{th}}} \right)} \right]} \right\}\notag\\
& \hspace*{0.3cm}\times \left[ {1 - {\Xi _s}\gamma \left( {\frac{2}{\alpha },\frac{{\left( {{\rho _{GF}}{t_s}\left( {{x_s},0,{\tau _{th}}} \right) + 1} \right)\gamma _{th}^{GB}R_1^\alpha }}{{{\rho _{GB}}{\lambda _{GB}}}}} \right)} \right],
\end{align}
\emph{where ${\Xi _s} = \frac{2}{{\alpha R_1^2\gamma _{th}^{GB}}}{\left( {\frac{{{\rho _{GB}}{\lambda _{GB}}}}{{{\rho _{GF}}{t_s}\left( {{x_s},0,{\tau _{th}}} \right) + 1}}} \right)^{\frac{2}{\alpha }}}$}.
\begin{IEEEproof}
Substituting \eqref{pdf_h/d_far}, \eqref{CDF_near1} and \eqref{CDF_near2} into \eqref{T1}, the derivations can be obtained by Chebyshev-gauss quadrature.
\end{IEEEproof}
\end{corollary}

\begin{corollary}\label{COPGB2}
\emph{Assume that all of the GF users can access into the targeted GB channels, which means $g_{GF} \ll \tau_{th}$. In this case, we can derive the approximated expressions of OP when $\tau_{th}  \to \infty $ as}
\begin{align}
&P_{out,{p_1}}^{GB,I} = F_{{g_{GB,j}}}^{near}\left( {{\tau _{th}}} \right) - \frac{{2b_{1,2}^{GF}}}{{\alpha R_1^2}}{\left( {\frac{{{\rho _{GB}}{\lambda _{GB}}}}{{\gamma _{th}^{GB}{\rho _{GF}}}}} \right)^{\frac{2}{\alpha }}}\notag\\
& \hspace*{0.2cm}\times \left[ {U\left( {\frac{{\gamma _{th}^{GB}{\rho _{GF}}R_1^\alpha }}{{{\rho _{GB}}{\lambda _{GB}}}},b_{2,2}^{GF}} \right) - U\left( {\frac{{\gamma _{th}^{GB}{\rho _{GF}}R_1^\alpha }}{{{\rho _{GB}}{\lambda _{GB}}}},b_{2,1}^{GF}} \right)} \right]  ,
\end{align}
\emph{where $U\left( {a,t} \right)$ can be seen as \textbf{Corollary \ref{AOPGB21}}.}
\begin{IEEEproof}
Conditioned on $\tau_{th}  \to \infty $ and based on \eqref{pdf_h/d_far} and \eqref{CDF_near2}, \eqref{T1} can be derived by substituting Eq.[2.10.6.2] in \cite{Prudnikov2}.
\end{IEEEproof}
\end{corollary}

We then investigate the outage performance of the GF users. Based on the \textbf{Theorem \ref{OPGF1}}, we can derive the exact and approximated closed-form expressions in \textbf{Corollary \ref{COPGF1}} and \textbf{Corollary \ref{COPGF2}} with perfect SIC procedure.

\begin{theorem}\label{OPGF1}
\emph{Note that the GF users are decoded at the last decoding orders in Scenario I. Thus, two outage situations are involved: \textit{a)} the messages of the GB users cannot be detected so that SIC procedure is not successful and \textit{b)} the BS can detect the messages of the GB users but cannot detect that of the GF users. Based on \textbf{Theorem \ref{OPGB1}}, the first situation has been analyzed. Thus, we derive $Q_2$ in \eqref{P_out_s1_GF} as}
\begin{align}\label{T2}
{Q_2} &= \int_0^{\min \left( {\frac{{\gamma _{th}^{GF}}}{{{\rho _{GF}}}},{\tau _{th}}} \right)} {\left[ {1 - F_{{g_{GB,j}}}^{near}\left( {\frac{{\gamma _{th}^{GB}{\rho _{GF}}x + \gamma _{th}^{GB}}}{{{\rho _{GB}}}}} \right)} \right]} \notag\\
&\hspace*{0.3cm} \times f_{{g_{GF,i}}}^{far}\left( x \right)dx,
\end{align}
\emph{where \textbf{Corollary \ref{COPGF1}} and \textbf{Corollary \ref{COPGF2}} can provide two closed-from expressions. Thus, based on \textbf{Theorem \ref{OPGB1}} and \textbf{Theorem \ref{OPGF1}}, the final outage probability of GF users in Scenario I can be derived as $P_{out,{p_1}}^{GF,I} = {Q_2} + P_{out,{p_1}}^{GB,I}$.}
\end{theorem}

\begin{corollary}\label{COPGF1}
\emph{Based on Chebyshev-gauss quadrature, the closed-form expressions of $Q_2$ can be derived as}
\begin{align}
{Q_2}&=\sum\limits_{s = 1}^S {\frac{{{\omega _s}\min \left( {\frac{{\gamma _{th}^{GF}}}{{{\rho _{GF}}}},{\tau _{th}}} \right){{\left( {{\rho _{GB}}{\lambda _{GB}}} \right)}^{{b_3} - 1}}}}{{\alpha R_1^2{{\left( {1 - {\iota _{s,{\rm{0}}}}^2} \right)}^{ - \frac{{\rm{1}}}{{\rm{2}}}}}{{\left( {\gamma _{th}^{GB}{\rho _{GF}}x + \gamma _{th}^{GB}} \right)}^{{b_3} - 1}}}}} \notag\\
& \hspace*{0.3cm} \times \frac{{b_{1,2}^{GF}}}{{{{\iota _{s,{\rm{0}}}}}^{{b_3}}}}\gamma \left[ {{b_3} - 1,\frac{{b_{2,1}^{GB}\gamma _{th}^{GB}\left( {{\rho _{GF}}{{\iota _{s,{\rm{0}}}}} + 1} \right)}}{{{\rho _{GB}}}}} \right]\notag\\
&\hspace*{0.3cm} \times \left[ {\gamma \left( {{b_3},b_{2,2}^{GF}{{\iota _{s,{\rm{0}}}}}} \right) - \gamma \left( {{b_3},b_{2,1}^{GF}{{\iota _{s,{\rm{0}}}}}} \right)} \right],
\end{align}
\emph{where ${\iota _{s,{\rm{0}}}}{\rm{ = }}{t_s}\left( {{x_s},0,{\sigma _{\rm{0}}}} \right)$ and ${\sigma _0} = \min \left( {\frac{{\gamma _{th}^{GF}}}{{{\rho _{GF}}}},{\tau _{th}}} \right)$.}
\begin{IEEEproof}
See \textbf{Corollary \ref{COPGB1}}.
\end{IEEEproof}
\end{corollary}

\begin{corollary}\label{COPGF2}
\emph{Based on the expansions of lower incomplete gamma functions as $\gamma \left( {a,b} \right) = \sum\limits_{n=0}^\infty  {\frac{{{{\left( { - 1} \right)}^n}{b^{a + n}}}}{{n!\left( {a + n} \right)}}} $, the closed-form expressions of $Q_2$ can be equally derived as}
\begin{align}
{Q_2} &= {C_1}\sum\limits_{n = 0}^\infty  {\sum\limits_{t = 0}^n {\dbinom{n}{r}} } {\left( {\frac{{ - \gamma _{th}^{GB}{\rho _{GF}}}}{{{\rho _{GB}}}}} \right)^n}\frac{{{{\left( {b_{2,1}^{GB}} \right)}^{\frac{2}{\alpha } + n}}}}{{n!\left( {\frac{2}{\alpha } + n} \right)\rho _{GF}^t}}\notag\\
&\hspace*{0.3cm}  \times \left\{ {M\left[ {{\sigma _0},q,{b_3},b_{2,2}^{GF}} \right]} \right.\left. { - M\left[ {{\sigma _0},q,{b_3},b_{2,1}^{GF}} \right]} \right\} .
\end{align}
\begin{IEEEproof}
See \textbf{Corollary \ref{AOPGF21}}.
\end{IEEEproof}
\end{corollary}

\subsection{Asymptotic OP under Dynamic Protocol in Scenario I}
Diversity orders as intuitive indicators present performance changing with transmit SNR $\rho_G= P_{G}/\sigma^2$ and $G \in\{GF,GB\}$. When analyzing the asymptotic performance, the condition with high transmit powers of the GB users and fixed transmit powers of the GF users is assumed. Note that $P_{GB} \to \infty $ equals $\rho_{GB} \to \infty $. Additionally, due to the statistic of $g_{G,k}$ is the expressions with the lower incomplete gamma functions, we derive the asymptotic expressions by the expansions of the lower incomplete gamma functions remaining the first two items, denoted as
\begin{align}\label{gamma_asy}
\gamma \left( {a,b} \right) = \sum\limits_{n = 0}^\infty  {\frac{{{{\left( { - 1} \right)}^n}{b^{a + n}}}}{{n!\left( {a + n} \right)}}}  = \frac{{{b^a}}}{a} - \frac{{{b^{a + 1}}}}{{a + 1}}  .
\end{align}

In the following, the asymptotic OP expressions of the GB and the GF users under dynamic protocol are derived as \textbf{Corollary \ref{dOPGBA1}} and \textbf{Corollary \ref{dOPGFA1}}.

\begin{corollary}\label{dOPGBA1}
\emph{Based on \textbf{Theorem \ref{OPGB2}}, two situations as $\gamma _{th}^{GB} > 1$ and $\gamma _{th}^{GB} \le 1$ are involved in this corollary. Assumed the expansions of lower incomplete gamma function with two items, the expressions for the GB users under two situations can be derived in terms of asymptotic OP, respectively. }

\emph{Conditioned on $\gamma _{th}^{GB} > 1$, the asymptotic OP can be derived as }
\begin{align}\label{co31}
&P_{out,{p_2}}^{GB,I,\infty } = {C_1}{\left( {\frac{{{\rho _{GB}}}}{{{\rho _{GF}}}}} \right)^{{b_3} - 1}}\left[ {U\left( {\frac{{b_{2,1}^{GB}{\rho _{GF}}}}{{{\rho _{GB}}}},b_{2,2}^{GF}} \right)} \right.\notag\\
&\hspace*{0.5cm}\left. { - U\left( {\frac{{b_{2,1}^{GB}{\rho _{GF}}}}{{{\rho _{GB}}}},b_{2,1}^{GF}} \right)} \right] - 1 + P_{out,{p_1}}^{GB,I,\infty }\left( \infty  \right),
\end{align}
\emph{where $P_{out,{p_1}}^{GB,I,\infty }\left( \infty  \right)$ is as \eqref{P_inf} in \textbf{Proposition \ref{OPGBA12}}.}

\emph{Conditioned on $\gamma _{th}^{GB} \le 1$, the asymptotic OP can be calculated as }
\begin{align}\label{co32}
P_{out,{p_2}}^{GB,I,\infty } = P_{out,{p_1}}^{GB,I,\infty }\left( {{\sigma _1}} \right) - {I_5},
\end{align}
\emph{where $P_{out,{p_1}}^{GB,I,\infty }( {\cdot})$ is expressed by \eqref{co1} and $I_5$ is as}
\begin{align}\label{I5}
{I_5} &= F_{{g_{GF}}}^{far}\left( {{\sigma _1}} \right) - \sum\limits_{s = 1}^S {{\Lambda _2}\left( {{\sigma _1},{\iota _{s,1}}} \right)} \gamma \left( {{b_3} - 1,b_{2,1}^{GB}\frac{{{\rho _{GF}}}}{{{\rho _{GB}}}}{\iota _{s,1}}} \right)\notag\\
 &\hspace*{0.3cm}\times \left[ {\gamma \left( {{b_3},b_{2,2}^{GF}{\iota _{s,1}}} \right) - \gamma \left( {{b_3},b_{2,1}^{GF}{\iota _{s,1}}} \right)} \right].
\end{align}
\begin{IEEEproof}
When $\gamma _{th}^{GB} > 1$, substituting \eqref{pdf_h/d_near}, \eqref{CDF_near1} and \eqref{gamma_asy} into the expressions of OP for the GB users, the asymptotic expressions can be obtained by utilizing Eq. [2.10.6.2] in \cite{Prudnikov2}.

When $\gamma _{th}^{GB} \le 1$, the derivations in \textbf{Proposition \ref{OPGBA12}} and Chebushev-gauss quadrature are invoked to carry out the final expressions.
\end{IEEEproof}
\end{corollary}

\begin{corollary}\label{dOPGFA1}
\emph{Conditioned that two outage situations are considered: \textit{a)} the outage thresholds of the GB users are lower than that of the GF users with the condition as $\gamma _{th}^{GB} > {{\gamma _{th}^{GF}} \mathord{\left/
 {\vphantom {{\gamma _{th}^{GF}} {\left( {1 + \gamma _{th}^{GF}} \right)}}} \right.
 \kern-\nulldelimiterspace} {\left( {1 + \gamma _{th}^{GF}} \right)}}$ or \textit{b)} the GB users have high outage thresholds as $\gamma _{th}^{GB} \le {{\gamma _{th}^{GF}} \mathord{\left/
 {\vphantom {{\gamma _{th}^{GF}} {\left( {1 + \gamma _{th}^{GF}} \right)}}} \right.
 \kern-\nulldelimiterspace} {\left( {1 + \gamma _{th}^{GF}} \right)}}$. Thus, using the asymptotic expressions of the lower incomplete gamma function, the asymptotic OP of the GF users can be derived respectively.}

 \emph{When $\gamma _{th}^{GB} > {{\gamma _{th}^{GF}} \mathord{\left/
 {\vphantom {{\gamma _{th}^{GF}} {\left( {1 + \gamma _{th}^{GF}} \right)}}} \right.
 \kern-\nulldelimiterspace} {\left( {1 + \gamma _{th}^{GF}} \right)}}$, the asymptotic expressions can be derived as}
\begin{align}\label{co4}
P_{out,{p_2}}^{GF,I,\infty }{\rm{ = }}F_{{g_{GF}}}^{far}\left( {{\sigma _{\rm{2}}}} \right) - P_{out,{p_1}}^{GB,I,\infty }\left( {{\sigma _{\rm{2}}}} \right){\rm{ + }}P_{out,{p_2}}^{GB,I}.
\end{align}

\emph{When $\gamma _{th}^{GB} \le {{\gamma _{th}^{GF}} \mathord{\left/
 {\vphantom {{\gamma _{th}^{GF}} {\left( {1 + \gamma _{th}^{GF}} \right)}}} \right.
 \kern-\nulldelimiterspace} {\left( {1 + \gamma _{th}^{GF}} \right)}}$, we can express the asymptotic expressions as}
 \begin{align}\label{co4}
&P_{out,{p_2}}^{GF,I,\infty }=F_{{g_{GF}}}^{far}\left( {{\sigma _{\rm{1}}}} \right) - P_{out,{p_1}}^{GB,I,\infty }\left( {{\sigma _{\rm{1}}}} \right){\rm{ + }}P_{out,{p_2}}^{GB,I}\notag\\
&\hspace*{0.5cm}{\rm{ + }}{\Lambda _2}\left[ {({\sigma _2} - {\sigma _1}),{\iota _{s,12}}} \right]\gamma \left( {{b_3} - 1,\frac{{b_{2,1}^{GB}{\rho _{GF}}}}{{{\rho _{GB}}}}{\iota _{s,12}}} \right)\notag\\
&\hspace*{0.5cm} \times \left[ {\gamma \left( {{b_3},b_{2,2}^{GF}{\iota _{s,12}}} \right) - \gamma \left( {{b_3},b_{2,1}^{GF}{\iota _{s,12}}} \right)} \right].
\end{align}

\begin{IEEEproof}
Based on the derivations of \eqref{CDF_far1}, \eqref{P_inf}, \eqref{co31} and \eqref{co32}, the asymptotic expressions of OP for the GF users can be derived.
\end{IEEEproof}
\end{corollary}

\subsection{Asymptotic OP under Open-loop Protocol in Scenario I }
The condition of open-loop protocol is $P_{GF}g_{GF} < \tau_{th}$. Note that performance for a certain network with a higher diversity order outperforms. When $P_{GB} \to \infty $ , diversity orders under Scenario I are analyzed as \textbf{Corollary \ref{OPGBA1}} and \textbf{Corollary \ref{OPGFA1}} to investigate the outage performance in high SNR region.

\begin{corollary}\label{OPGBA1}
\emph{Conditioned that only the transmit powers of the GB users are ultra-high, denoted as $P_{GB} \to \infty$, whereas the transmit power of the GF users $P_{GF} $ are fixed, we can carry out the high SNR expressions of the GB users in terms of OP. Thus, in Scenario I under open-loop protocol, the asymptotic expressions of OP for the GB users can be derived as}
\begin{align}\label{co1}
&P_{out,{p_1}}^{GB,I,\infty }(\tau_{th}) = {\Upsilon }\left[ {{\rho _{GF}}b_{1,2}^{GF}M\left( {{\tau _{th}},2 - {b_3},{b_3},b_{2,2}^{GF}} \right)} \right.\notag\\
&\hspace*{0.3cm} \left. { - {\rho _{GF}}b_{1,2}^{GF}M\left( {{\tau _{th}},2 - {b_3},{b_3},b_{2,1}^{GF}} \right){\rm{ + }}F_{{g_{GF}}}^{far}\left( {{\tau _{th}}} \right)} \right],
\end{align}
\emph{where ${\Upsilon } = \frac{{{\rm{2}}{{\left( {b_{2,1}^{GB}} \right)}^{\frac{2}{\alpha }{\rm{ + 1}}}}{\lambda _{GB}}^{\frac{2}{\alpha }}\gamma _{th}^{GB}}}{{\left( {\alpha {\rm{ + 2}}} \right)R_1^2{\rho _{GB}}}}$.}
\begin{IEEEproof}
Substituting \eqref{gamma_asy} into the expressions of OP of the GB users under open-loop protocol, it can be derived as
\begin{align}
P_{out,{p_1}}^{GB,I,\infty }(\tau_{th}) = \int_0^{{\tau _{th}}} {\left[ {{\Upsilon }\left( {{\rho _{GF}}x + {\rm{1}}} \right)} \right]f_{{g_{GF}}}^{far}\left( x \right)} dx.
\end{align}

Utilizing Eq. [2.10.2.2] in \cite{Prudnikov2}, the final closed-form expressions can be obtained.
\end{IEEEproof}
\end{corollary}

\begin{proposition}\label{OPGBA12}
\emph{One special case is described that all the GF users can access into the GB channels when $g_{GF} \ll \tau_{th}$. Thus, we can assume $\tau_{th} \to \infty $ to derive the approximated expressions of OP for the GB users as}
\begin{align}\label{P_inf}
&P_{out,{p_1}}^{GB,I,\infty }(\infty){\rm{ = }}\notag\\
&{\Upsilon }\left\{ {\frac{{{\rho _{GF}}b_{{\rm{1}},{\rm{2}}}^{GF}\Gamma \left( {\rm{2}} \right)}}{{\left( {2 - {b_3}} \right)}}\left[ {{{\left( {b_{2,1}^{GF}} \right)}^{{b_3} - 2}} - {{\left( {b_{2,2}^{GF}} \right)}^{{b_3} - 2}}} \right]{\rm{ + 1}}} \right\}.
\end{align}
\begin{IEEEproof}
Based on \textbf{Corollary \ref{OPGBA1}} and Eq. [2.10.2.2] in \cite{Prudnikov2}, this proposition can be proved.
\end{IEEEproof}
\end{proposition}

\begin{corollary}\label{OPGFA1}
\emph{Note that same assumptions in \textbf{Corollary \ref{OPGBA1}} are proposed in this corollary. Thus, the asymptotic expressions of OP for the GF users can be derived as}
\begin{align}\label{co2}
&P_{out,{p_1}}^{GF,I,\infty } = F_{{g_{GF}}}^{far}\left[ \min \left( {{\sigma _2},{\tau _{th}}} \right)\right]\notag\\
 &\hspace*{0.3cm} - P_{out,{p_1}}^{GB,I,\infty }\left[ \min \left( {{\sigma _2},{\tau _{th}}} \right) \right] + P_{out,{p_1}}^{GB,I,\infty }(\tau_{th}),
\end{align}
{where the final expressions can be obtained by substituting \eqref{CDF_far1} and \eqref{co1}.}
\begin{IEEEproof}
Substituting the expansions of lower incomplete gamma function into the expressions of OP for the GF users, one can be achieved as
\begin{align}
P_{out,{p_1}}^{GF,I,\infty } &= \int_0^{\min \left( {{\sigma _2},{\tau _{th}}} \right)} {\left[ {1 - {\Upsilon}\left( {{\rho _{GF}}x + {\rm{1}}} \right)} \right]f_{{g_{GF}}}^{far}\left( x \right)} dx\notag\\
 &\hspace*{0.3cm} + P_{out,{p_1}}^{GB,I,\infty }\left( {{\tau _{th}}} \right).
\end{align}

Based on the derivations in \textbf{Corollary \ref{OPGBA1}} and change the upper limits from $\tau_{th} $ to $\min \left( {{\sigma _2},{\tau _{th}}} \right)$, the asymptotic expressions can be derived as \eqref{co2}.
\end{IEEEproof}
\end{corollary}


\begin{remark}\label{D_GB}
The expressions of diversity orders for the GB users can be presented as:
\begin{align}\label{dogb}
{d_{GB}} =  - \mathop {\lim }\limits_{{\rho_{GB} } \to \infty } \frac{{\log {P^{GB,I,\infty}_{out,p_a}(\rho_{GB})}}}{{\log {\rho_{GB} }}}=1,
\end{align}
where $P^{GB,I,\infty}_{out,p_a}$ is the asymptotic OP in Scenario I with $a \in \{1,2\}$ as different protocols and $\rho_{GB}$ means the transmit SNR of the GB users.
\end{remark}

\begin{remark}\label{D_GF}
We can define the expressions of diversity orders for the GF user as:
\begin{align}\label{dogf}
{d_{GF}} =  - \mathop {\lim }\limits_{{\rho_{GF} } \to \infty } \frac{{\log {P^{GF,I,\infty }_{out,p_a}(\rho_{GF})}}}{{\log {\rho_{GF}}}}=0,
\end{align}
where $P^{GF,I,\infty }_{out,p_a}$ is the asymptotic OP and $\rho_{GF}$ means the transmit SNR for the GF users..
\end{remark}

\section{Outage Performance in Scenario II }

In Scenario II, the GB users are arranged into the ring as the far users while the GF users are settled in the disc as near users. Thus, the SIC orders can be determined by the distance of users shown as: the GF users at the first stage of SIC orders and the GB users at last. Based on the aforementioned two protocols, i.e., open-loop protocol and dynamic protocol, we derive the exact expressions of OP as \textbf{Theorem \ref{GF_p2_T2}} to \textbf{Theorem \ref{GB_p1_T2}} with closed-form expressions via several corollaries.

\subsection{Analytical OP under Dynamic Protocol in Scenario II}

\begin{theorem}\label{GF_p2_T2}
\emph{Conditioned on $\gamma _{th}^{GB} < 1$ and $\gamma _{th}^{GB} \ge 1$, the derivations can be derived by various expressions. }

\emph{\textit{a)} When $\gamma _{th}^{GB} \ge 1$, the expressions of OP for the GF users can be obtained as}
\begin{align}
P_{out,{p_2}}^{GF,II}& = \underbrace {\int_0^\infty  {F_{{g_{GF}}}^{near}\left( {\frac{{\gamma _{th}^{GF}{\rho _{GB}}}}{{{\rho _{GF}}}}x + \frac{{\gamma _{th}^{GF}}}{{{\rho _{GF}}}}} \right)} f_{{g_{GB}}}^{far}\left( x \right)dx}_{{I_5}}\notag\\
&\hspace*{0.3cm} - \underbrace {\int_0^\infty  {F_{{g_{GF}}}^{near}\left( {\frac{{{\rho _{GB}}}}{{{\rho _{GF}}}}x} \right)} f_{{g_{GB}}}^{far}\left( x \right)dx}_{{I_6}}.
\end{align}

\emph{\textit{b)} When $\gamma _{th}^{GB} < 1$, we can write the expressions of OP for the GF users as   }
\begin{align}
P_{out,{p_2}}^{GF,II} &= \int_0^{{\sigma _3}} {F_{{g_{GF}}}^{near}\left( {\frac{{\gamma _{th}^{GF}{\rho _{GB}}}}{{{\rho _{GF}}}}x + \frac{{\gamma _{th}^{GF}}}{{{\rho _{GF}}}}} \right)} f_{{g_{GB}}}^{far}\left( x \right)dx\notag\\
&\hspace*{0.3cm} - \int_0^{{\sigma _3}} {F_{{g_{GF}}}^{near}\left( {\frac{{{\rho _{GB}}}}{{{\rho _{GF}}}}x} \right)} f_{{g_{GB}}}^{far}\left( x \right)dx,
\end{align}
\emph{where ${\sigma _3} = {{\gamma _{th}^{GF}} \mathord{\left/
 {\vphantom {{\gamma _{th}^{GF}} {\left( {{\rho _{GF}} - {\rho _{GB}}\gamma _{th}^{GF}} \right)}}} \right.
 \kern-\nulldelimiterspace} {\left( {{\rho _{GF}} - {\rho _{GB}}\gamma _{th}^{GF}} \right)}}$ and the closed-form expressions are given in \textbf{Corollary \ref{16}}.}
\end{theorem}

\begin{corollary}\label{16}
\emph{We assume the transmit SNR of the GF users as $\rho_{GF} \to \infty$ for the case of $\gamma _{th}^{GB} \ge 1$, which means the GF users have satisfying channel conditions. Thus, one approximated expression of $I_5$ and the accurate derivation of $I_6$ can be calculated as }
\begin{align}
{I_5} &= 1 - {C_2}{\left( {{\rho _{GB}}{\sigma _2}} \right)^{1 - {b_3}}}U\left( {b_{2,1}^{GF}{\rho _{GB}}{\sigma _2},b_{2,2}^{GB}} \right)\notag\\
&\hspace*{0.3cm} - {C_2}{\left( {{\rho _{GB}}{\sigma _2}} \right)^{1 - {b_3}}}U\left( {b_{2,1}^{GF}{\rho _{GB}}{\sigma _2},b_{2,1}^{GB}} \right)
\end{align}
\emph{and }
\begin{align}\label{I6}
{I_6}& = 1 - {C_2}{\left( {\frac{{{\rho _{GB}}}}{{{\rho _{GF}}}}} \right)^{{\rm{1}} - {b_3}}}U\left( {\frac{{b_{2,1}^{GF}{\rho _{GB}}}}{{{\rho _{GF}}}},b_{2,2}^{GB}} \right)\notag\\
&\hspace*{0.3cm} - {C_2}{\left( {\frac{{{\rho _{GB}}}}{{{\rho _{GF}}}}} \right)^{{\rm{1}} - {b_3}}}U\left( {\frac{{b_{2,1}^{GF}{\rho _{GB}}}}{{{\rho _{GF}}}},b_{2,1}^{GB}} \right),
\end{align}
\emph{where ${C_2} = {{2b_{1,2}^{GB}\lambda _{{\rm{GF}}}^{{b_3} - 1}} \mathord{\left/
 {\vphantom {{2b_{1,2}^{GB}\lambda _{{\rm{GF}}}^{{b_3} - 1}} {\left( {\alpha R_1^2} \right)}}} \right.
 \kern-\nulldelimiterspace} {\left( {\alpha R_1^2} \right)}}$. Substituting the expressions of $I_5$ and $I_6$, the closed-form expressions of OP for the GF users can be obtained.}

\emph{Conditioned on $\gamma _{th}^{GB} < 1$, based on Chebyshev-Gauss quadrature, the OP for the GF users can be figured out as }
\begin{align}
&P_{out,{p_2}}^{GF,II} = \sum\limits_{s = 1}^S {{\Lambda _3}\left( {{\sigma _{\rm{3}}},{\iota _{s,{\rm{3}}}}} \right)\gamma \left( {{b_3} - 1,b_{2,1}^{GF}\frac{{{\rho _{GB}}}}{{{\rho _{GF}}}}{\iota _{s,{\rm{3}}}}} \right)}\notag \\
 &\hspace*{0.3cm}\times \left[ {\gamma \left( {{b_3},b_{2,2}^{GB}{\iota _{s,{\rm{3}}}}} \right) - \gamma \left( {{b_3},b_{2,1}^{GB}{\iota _{s,{\rm{3}}}}} \right)} \right]\notag\\
 &\hspace*{0.3cm}- {\Lambda _{\rm{3}}}\left( {{\sigma _3},{\iota _{s,{\rm{3}}}}} \right)\gamma \left[ {{b_3} - 1,b_{2,1}^{GF}{\sigma _2}\left( {{\rho _{GB}}{\iota _{s,{\rm{3}}}} + 1} \right)} \right]\notag\\
&\hspace*{0.3cm} \times \left[ {\gamma \left( {{b_3},b_{2,2}^{GB}{\iota _{s,{\rm{3}}}}} \right) - \gamma \left( {{b_3},b_{2,1}^{GB}{\iota _{s,{\rm{3}}}}} \right)} \right],
\end{align}
\emph{where ${\Lambda _3}\left( {a,x} \right) = \frac{{{C_2}}}{2}a{\omega _s}{x^{1 - 2{b_3}}}{\left( {1 - {x^2}} \right)^{\frac{{\rm{1}}}{{\rm{2}}}}}\rho _{GF}^{{b_3} - 1}\rho _{GB}^{1 - {b_3}}$ and ${\iota _{s,3}} = {t_s}\left( {{x_s},0,{\sigma _3}} \right)$}
\end{corollary}

\begin{theorem}\label{GB_p2_T2}
\emph{Under two situations as $\gamma _{th}^{GF} > {{\gamma _{th}^{GB}} \mathord{\left/
 {\vphantom {{\gamma _{th}^{GB}} {\left( {\gamma _{th}^{GB} + 1} \right)}}} \right.
 \kern-\nulldelimiterspace} {\left( {\gamma _{th}^{GB} + 1} \right)}}$ and $\gamma _{th}^{GF} \le {{\gamma _{th}^{GB}} \mathord{\left/
 {\vphantom {{\gamma _{th}^{GB}} {\left( {\gamma _{th}^{GB} + 1} \right)}}} \right.
 \kern-\nulldelimiterspace} {\left( {\gamma _{th}^{GB} + 1} \right)}}$, the closed-form OP expressions of the GB users can be derived.}

\emph{\textit{a)} When $\gamma _{th}^{GF} > {{\gamma _{th}^{GB}} \mathord{\left/
 {\vphantom {{\gamma _{th}^{GB}} {\left( {\gamma _{th}^{GB} + 1} \right)}}} \right.
 \kern-\nulldelimiterspace} {\left( {\gamma _{th}^{GB} + 1} \right)}}$, the OP expressions are derived as}
\begin{align}
&P_{out,{p_2}}^{GB,II} = \sum\limits_{s = 1}^S {{\Lambda _4}\left( {{\sigma _3},{\iota _{s,3}}} \right)\gamma \left[ {{b_3} - 1,b_{2,1}^{GF}{\sigma _2}\left( {{\rho _{GB}}{\iota _{s,3}} + 1} \right)} \right]} \notag\\
 &\hspace*{0.3cm}\times \left[ {\gamma \left( {{b_3},b_{2,2}^{GB}{\iota _{s,3}}} \right) - \gamma \left( {{b_3},b_{2,1}^{GB}{\iota _{s,3}}} \right)} \right] + P_{out,{p_2}}^{GF,II}.
\end{align}

 \emph{\textit{b)} When $\gamma _{th}^{GF} \le {{\gamma _{th}^{GB}} \mathord{\left/
 {\vphantom {{\gamma _{th}^{GB}} {\left( {\gamma _{th}^{GB} + 1} \right)}}} \right.
 \kern-\nulldelimiterspace} {\left( {\gamma _{th}^{GB} + 1} \right)}}$, we can obtain the closed-form expressions as}
\begin{align}
&P_{out,{p_2}}^{GB,II} = \sum\limits_{s = 1}^S {{\Lambda _4}\left( {{\sigma _3},{\iota _{s,3}}} \right)\gamma \left[ {{b_3} - 1,b_{2,1}^{GF}{\sigma _2}\left( {{\rho _{GB}}{\iota _{s,3}} + 1} \right)} \right]} \notag\\
 &\hspace*{0.3cm}\times \left[ {\gamma \left( {{b_3},b_{2,2}^{GB}{\iota _{s,3}}} \right) - \gamma \left( {{b_3},b_{2,1}^{GB}{\iota _{s,3}}} \right)} \right]\notag\\
 &\hspace*{0.3cm}+ {\Lambda _3}\left[ {({\sigma _3} - {\sigma _4}),{\iota _{s,34}}} \right]\gamma \left( {{b_3} - 1,b_{2,1}^{GF}{\sigma _2}{\iota _{s,34}}} \right)\notag\\
 &\hspace*{0.3cm}\times \left[ {\gamma \left( {{b_3},b_{2,2}^{GB}{\iota _{s,34}}} \right) - \gamma \left( {{b_3},b_{2,1}^{GB}{\iota _{s,34}}} \right)} \right] + P_{out,{p_2}}^{GF,II},
\end{align}
\emph{where ${\sigma _4} = {{\gamma _{th}^{GB}} \mathord{\left/
 {\vphantom {{\gamma _{th}^{GB}} {{\rho _{GB}}}}} \right.
 \kern-\nulldelimiterspace} {{\rho _{GB}}}}$, ${\iota _{s,34}} = {t_s}\left( {{x_s},{\sigma _3},{\sigma _4}} \right)$ and ${\Lambda _4}(a,x) = \frac{{{C_2}}}{2}{x^{ - {b_3}}}{\omega _s}a{\left( {1 - {x^2}} \right)^{\frac{{\rm{1}}}{{\rm{2}}}}}{\left( {\gamma _{th}^{GF}} \right)^{1 - {b_3}}}\rho _{GF}^{{b_3} - 1}$ $\times {\left( {{\rho _{GB}}x + 1} \right)^{1 - {b_3}}}$.}
\end{theorem}

\subsection{Analytical OP under Open-loop Protocol in Scenario II}

\begin{theorem}\label{GF_p1_T2}
\emph{Conditioned on Scenario II with open-loop protocol, the derivations of OP for the GF users vary on two situations, which are expressed that: \textit{a)} the first situation is when the transmit SNR of the GF users is high enough to meet ${\rho _{GF}} \ge {{\gamma _{th}^{GF}} \mathord{\left/
 {\vphantom {{\gamma _{th}^{GF}} {{\tau _{th}}}}} \right.
 \kern-\nulldelimiterspace} {{\tau _{th}}}}$ and \textit{b)} another is when ${\rho _{GF}} < {{\gamma _{th}^{GF}} \mathord{\left/
 {\vphantom {{\gamma _{th}^{GF}} {{\tau _{th}}}}} \right.
 \kern-\nulldelimiterspace} {{\tau _{th}}}}$.}
 \emph{Thus, when ${\rho _{GF}} \ge {{\gamma _{th}^{GF}} \mathord{\left/
 {\vphantom {{\gamma _{th}^{GF}} {{\tau _{th}}}}} \right.
 \kern-\nulldelimiterspace} {{\tau _{th}}}}$, the OP expressions of the GF users can be derived as}
\begin{align}
P_{out,{p_1}}^{GF,II} &= \int_{\sigma _5^{ - 1}}^\infty  {F_{{g_{GF}}}^{near}\left( {\frac{{{\rho _{GB}}\gamma _{th}^{GF}}}{{{\rho _{GF}}}}x + \frac{{\gamma _{th}^{GF}}}{{{\rho _{GF}}}}} \right)f_{{g_{GB}}}^{far}\left( x \right)dx} \notag\\
 &\hspace*{0.3cm}- \int_{\sigma _5^{ - 1}}^\infty  {F_{{g_{GF}}}^{near}\left( {{\tau _{th}}} \right)f_{{g_{GB}}}^{far}\left( x \right)d} x,
\end{align}
\emph{and when ${\rho _{GF}} < {{\gamma _{th}^{GF}} \mathord{\left/
 {\vphantom {{\gamma _{th}^{GF}} {{\tau _{th}}}}} \right.
 \kern-\nulldelimiterspace} {{\tau _{th}}}}$, we can obtain the OP expressions for the GF users as }
\begin{align}
P_{out,{p_1}}^{GF,II}& = \int_0^\infty  {F_{{g_{GF}}}^{near}\left( {\frac{{{\rho _{GB}}\gamma _{th}^{GF}}}{{{\rho _{GF}}}}x + \frac{{\gamma _{th}^{GF}}}{{{\rho _{GF}}}}} \right)f_{{g_{GB}}}^{far}\left( x \right)dx} \notag\\
&\hspace*{0.3cm}- \int_0^\infty  {F_{{g_{GF}}}^{near}\left( {{\tau _{th}}} \right)f_{{g_{GB}}}^{far}\left( x \right)d} x,
\end{align}
\emph{where ${\sigma _5} = {{\gamma _{th}^{GF}{\rho _{GB}}} \mathord{\left/
 {\vphantom {{\gamma _{th}^{GF}{\rho _{GB}}} {\left( {{\rho _{GF}}{\tau _{th}} - \gamma _{th}^{GF}} \right)}}} \right.
 \kern-\nulldelimiterspace} {\left( {{\rho _{GF}}{\tau _{th}} - \gamma _{th}^{GF}} \right)}}$. \textbf{Corollary \ref{co17}} and \textbf{Corollary \ref{co18}} show the closed-form expressions of OP.}
\end{theorem}

\begin{corollary}\label{co17}
\emph{Conditioned on low channel gain thresholds and high transmit SNR of the GF users, denoted as $\tau_{th} \to 0$ and $\rho_{GF} \to \infty$, which means the special case that: \textit{a)} all the GF users can join into the channels occupied by the GB users and \textit{b)} the GF users experience good channel conditions. Under the situation ${\rho _{GF}} < {{\gamma _{th}^{GF}} \mathord{\left/
 {\vphantom {{\gamma _{th}^{GF}} {{\tau _{th}}}}} \right.
 \kern-\nulldelimiterspace} {{\tau _{th}}}}$, the OP of the GF users can be approximated as}
\begin{align}
&P_{out,{p_1}}^{GF,II}{\rm{ = 1}} - F_{{g_{GF}}}^{near}\left( {{\tau _{th}}} \right) - {C_2}{\left( {\frac{{{\rho _{GF}}}}{{{\rho _{GB}}\gamma _{th}^{GF}}}} \right)^{{b_3} - 1}}\notag\\
& \times \left[ {U\left( {\frac{{b_{2,1}^{GF}{\rho _{GB}}\gamma _{th}^{GF}}}{{{\rho _{GF}}}},b_{2,2}^{GB}} \right) - U\left( {\frac{{b_{2,1}^{GF}{\rho _{GB}}\gamma _{th}^{GF}}}{{{\rho _{GF}}}},b_{2,1}^{GB}} \right)} \right].
\end{align}
\begin{IEEEproof}
\emph{Based on \textbf{Lemma \ref{lemma:h/d}} and \textbf{Corollary \ref{COPGB2}}, we can proof this corollary.}
\end{IEEEproof}
\end{corollary}

\begin{corollary}\label{co18}
\emph{Conditioned on ${\rho _{GF}} \ge {{\gamma _{th}^{GF}} \mathord{\left/
 {\vphantom {{\gamma _{th}^{GF}} {{\tau _{th}}}}} \right.
 \kern-\nulldelimiterspace} {{\tau _{th}}}}$, we can derive the closed-form expressions of OP of the GF users as }
\begin{align}
P_{out,{p_1}}^{GF,II} &= \sum\limits_{s = 1}^S {\left\{ {{\vartheta _1}\gamma \left( {{b_3} - 1,b_{2,1}^{GF}{\tau _{th}}} \right)} \right.}\notag \\
 &\hspace*{0.3cm}- \left. {{\vartheta _2}\gamma \left[ {{b_3} - 1,b_{2,1}^{GF}\gamma _{th}^{GF}\left( {\frac{{{\rho _{GB}} + {\iota _{s,5}}}}{{{\rho _{GF}}{\iota _{s,5}}}}} \right)} \right]} \right\}\notag\\
&\hspace*{0.3cm} \times \left[ {\gamma \left( {{b_3},\frac{{b_{2,2}^{GB}}}{{{\iota _{s,5}}}}} \right) - \gamma \left( {{b_3},\frac{{b_{2,1}^{GB}}}{{{\iota _{s,5}}}}} \right)} \right],
\end{align}
\emph{where ${\iota _{s,5}} = {t_s}\left( {{x_s},0,{\sigma _5}} \right)$, ${\vartheta _1} = \tau _{th}^{1 - {b_3}}{\Lambda _5}\left( {{\sigma _5},{\iota _{s,5}}} \right)$, ${\vartheta _2} = {\left( {{\rho _{GF}}{\iota _{s,5}}} \right)^{{b_3} - 1}}{\left( {{\rho _{GB}}\gamma _{th}^{GF} + \gamma _{th}^{GF}{\iota _{s,5}}} \right)^{1 - {b_3}}}{\Lambda _5}\left( {{\sigma _5},{\iota _{s,5}}} \right)$ and ${\Lambda _5}\left( {a,x} \right) = \frac{{{C_2}}}{2}{a^{ - 1}}{\omega _s}{x^{{b_3} - 2}}{\left( {1 - {x^2}} \right)^{\frac{{\rm{1}}}{{\rm{2}}}}}$. }
\end{corollary}

\begin{theorem}\label{GB_p1_T2}
\emph{After SIC procedure, the OP expressions of the GB users can be written as $P_{out,{p_1}}^{GB,II} = {Q_3} + P_{out,{p_1}}^{GF,II}$, where $Q_3$ can be expressed as ${Q_3} = \Pr \left\{ {{g_{GF,i}} > \frac{{{\rho _{GB}}\gamma _{th}^{GF}}}{{{\rho _{GF}}}}{g_{GB,j}} + \frac{{\gamma _{th}^{GF}}}{{{\rho _{GF}}}},{g_{GB}} < \frac{{\gamma _{th}^{GB}}}{{{\rho _{GB}}}},{g_{GF}} > {\tau _{th}}} \right\}$. Based on \textbf{Theorem \ref{GF_p1_T2}} and \textbf{Corollary \ref{co19}}, we can achieve the closed-form OP expressions for the GB users under open-loop protocol in Scenario II.}
\end{theorem}

\begin{corollary}\label{co19}
\emph{With various range of the transmit SNR of the GF users, i.e., ${\rho _{GF}} \le {{\gamma _{th}^{GF}} \mathord{\left/
 {\vphantom {{\gamma _{th}^{GF}} {{\tau _{th}}}}} \right.
 \kern-\nulldelimiterspace} {{\tau _{th}}}}$, ${{\gamma _{th}^{GF}} \mathord{\left/
 {\vphantom {{\gamma _{th}^{GF}} {{\tau _{th}}}}} \right.
 \kern-\nulldelimiterspace} {{\tau _{th}}}} < {\rho _{GF}} < {{\gamma _{th}^{GF}\left( {1 + \gamma _{th}^{GB}} \right)} \mathord{\left/
 {\vphantom {{\gamma _{th}^{GF}\left( {1 + \gamma _{th}^{GB}} \right)} {{\tau _{th}}}}} \right.
 \kern-\nulldelimiterspace} {{\tau _{th}}}}$ and ${\rho _{GF}} \ge {{\gamma _{th}^{GF}\left( {1 + \gamma _{th}^{GB}} \right)} \mathord{\left/
 {\vphantom {{\gamma _{th}^{GF}\left( {1 + \gamma _{th}^{GB}} \right)} {{\tau _{th}}}}} \right.
 \kern-\nulldelimiterspace} {{\tau _{th}}}}$, we can obtain different derivations of OP for $Q_3$. }

 \emph{\textit{a)} When ${\rho _{GF}} \le {{\gamma _{th}^{GF}} \mathord{\left/
 {\vphantom {{\gamma _{th}^{GF}} {{\tau _{th}}}}} \right.
 \kern-\nulldelimiterspace} {{\tau _{th}}}}$, $Q_3$ can be derived as}
\begin{align}
{Q_3} &= \sum\limits_{s = 1}^S {{\Lambda _4}({\sigma _2},{\iota _{s,2}})\gamma \left[ {{b_3} - 1,\frac{{b_{2,1}^{GF}\gamma _{th}^{GF}}}{{{\rho _{GF}}}}\left( {{\rho _{GB}}{\iota _{s,2}} + 1} \right)} \right]} \notag\\
&\hspace*{0.3cm} \times \left[ {\gamma \left( {{b_3},b_{2,2}^{GB}{\iota _{s,2}}} \right) - \gamma \left( {{b_3},b_{2,1}^{GB}{\iota _{s,2}}} \right)} \right].
\end{align}

\emph{\textit{b)} When ${{\gamma _{th}^{GF}} \mathord{\left/
 {\vphantom {{\gamma _{th}^{GF}} {{\tau _{th}}}}} \right.
 \kern-\nulldelimiterspace} {{\tau _{th}}}} < {\rho _{GF}} < {{\gamma _{th}^{GF}\left( {1 + \gamma _{th}^{GB}} \right)} \mathord{\left/
 {\vphantom {{\gamma _{th}^{GF}\left( {1 + \gamma _{th}^{GB}} \right)} {{\tau _{th}}}}} \right.
 \kern-\nulldelimiterspace} {{\tau _{th}}}}$, the expressions of $Q_3$ can be derived as}
\begin{align}
{Q_3} &= F_{{g_{GB}}}^{far}\left( {\sigma _5^{ - 1}} \right)\left[ {1 - F_{{g_{GF}}}^{near}\left( {{\tau _{th}}} \right)} \right]{\rm{ + }}\sum\limits_{s = 1}^S {{\Lambda _4}(\varepsilon ,{\iota _{s,52}})} \notag\\
 &\hspace*{0.3cm}\times \gamma \left[ {{b_3} - 1,b_{2,1}^{GF}{\sigma _2}\left( {{\rho _{GB}}{\iota _{s,52}} + 1} \right)} \right]\notag\\
&\hspace*{0.3cm} \times \left[ {\gamma \left( {{b_3},b_{2,2}^{GB}{\iota _{s,52}}} \right) - \gamma \left( {{b_3},b_{2,1}^{GB}{\iota _{s,52}}} \right)} \right],
\end{align}
\emph{where ${\iota _{s,52}} = {t_s}\left( {{x_s},\sigma _5^{ - 1},{\sigma _2}} \right)$, $\varepsilon {\rm{ = }}{{\left( {\gamma _{th}^{GF}{\sigma _5} - {\rho _{GF}}} \right)} \mathord{\left/
 {\vphantom {{\left( {\gamma _{th}^{GF}{\sigma _5} - {\rho _{GF}}} \right)} {\left( {{\rho _{GF}}{\sigma _5}} \right)}}} \right.
 \kern-\nulldelimiterspace} {\left( {{\rho _{GF}}{\sigma _5}} \right)}}$. }

 \emph{\textit{c)} When ${\rho _{GF}} \ge {{\gamma _{th}^{GF}\left( {1 + \gamma _{th}^{GB}} \right)} \mathord{\left/
 {\vphantom {{\gamma _{th}^{GF}\left( {1 + \gamma _{th}^{GB}} \right)} {{\tau _{th}}}}} \right.
 \kern-\nulldelimiterspace} {{\tau _{th}}}}$, we can derive the expressions of $Q_3$ as }
\begin{align}
{Q_3} = F_{{g_{GB}}}^{far}\left( {\frac{{\gamma _{th}^{GF}}}{{{\rho _{GF}}}}} \right)\left[ {1 - F_{{g_{GF}}}^{near}\left( {{\tau _{th}}} \right)} \right].
\end{align}
\end{corollary}

\subsection{Asymptotic OP under Dynamic Protocol in Scenario II}

We assume that $P_{GF} \to \infty$ and a fixed $P_{GB}$ are presented to achieve the asymptotic expressions in Scenario II under dynamic protocol as \textbf{Corollary \ref{asy1}} and \textbf{Corollary \ref{co24}}.

\begin{corollary}\label{asy1}
\emph{Since $P_{GF} \to \infty$ means $\rho_{GF} \to \infty$, the asymptotic OP expressions of the GF users under dynamic protocol can be derived in the following two situations as:}
\emph{a) when $\gamma _{th}^{GB} > 1$,}
\begin{align}
P_{out,{p_2}}^{GF,II,\infty } = I_5^\infty  - {I_6},
\end{align}
\emph{where ${C_3} = {{2R_1^\alpha } \mathord{\left/
 {\vphantom {{2R_1^\alpha } {\left[ {\left( {\alpha  + 2} \right){\lambda _{GF}}} \right]}}} \right.
 \kern-\nulldelimiterspace} {\left[ {\left( {\alpha  + 2} \right){\lambda _{GF}}} \right]}}$ and $I_5^\infty $ can be expressed as}
\begin{align}
&I_5^\infty  = {C_3}{\sigma _2}b_{1,2}^{GB}\frac{{{\rho _{GB}}\Gamma \left( 2 \right)}}{{2 - {b_3}}}\left[ {{{\left( {b_{2,1}^{GB}} \right)}^{{b_3} - 2}} - {{\left( {b_{2,2}^{GB}} \right)}^{{b_3} - 2}}} \right]\notag\\
&\hspace*{0.3cm} + {C_3}{\sigma _2}b_{1,2}^{GB}\frac{{\Gamma \left( 1 \right)}}{{1 - {b_3}}}\left[ {{{\left( {b_{2,1}^{GB}} \right)}^{{b_3} - 1}} - {{\left( {b_{2,2}^{GB}} \right)}^{{b_3} - 1}}} \right],
\end{align}
\emph{b) and when $\gamma _{th}^{GB} \le 1$,}
\begin{align}
&P_{out,{p_2}}^{GF,II,\infty } =  - F_{{g_{GB}}}^{far}\left( {{\sigma _3}} \right) + \sum\limits_{s = 1}^S {\left[ {\frac{{{C_4}}}{{\iota _{s,{\rm{3}}}^{{b_3}}}}\left( {{\rho _{GB}}{\iota _{s,{\rm{3}}}} + 1} \right) + \frac{{{C_2}}}{{\iota _{s,{\rm{3}}}^{2{b_3} - 1}}}} \right.} \notag\\
 &\hspace*{0.8cm}\times \left. {{{\left( {\frac{{{\rho _{GF}}}}{{{\rho _{GB}}}}} \right)}^{{b_3} - 1}}\gamma \left( {{b_3} - 1,b_{2,1}^{GF}\frac{{{\rho _{GB}}}}{{{\rho _{GF}}}}x} \right)} \right]\notag\\
 &\hspace*{0.8cm}\times \left[ {\gamma \left( {{b_3},b_{2,2}^{GB}{\iota _{s,{\rm{3}}}}} \right) - \gamma \left( {{b_3},b_{2,1}^{GB}{\iota _{s,{\rm{3}}}}} \right)} \right],
\end{align}
\emph{where ${C_4} = {C_3}{\sigma _2}b_{1,2}^{GB}$.}
\end{corollary}

\begin{corollary}\label{co24}
\emph{In Scenario II with dynamic protocol, under the condition that $P_{GF} \to \infty$, we can derive the asymptotic expressions of OP for the GF users as the following two situations:}
\emph{a) when $\gamma _{th}^{GF} > {{\gamma _{th}^{GB}} \mathord{\left/
 {\vphantom {{\gamma _{th}^{GB}} {\left( {\gamma _{th}^{GB} + 1} \right)}}} \right.
 \kern-\nulldelimiterspace} {\left( {\gamma _{th}^{GB} + 1} \right)}}$,}
\begin{align}
P_{out,{p_2}}^{GB,II,\infty } = {Q_4}\left( {{\sigma _4 }} \right) + P_{out,{p_2}}^{GF,II,\infty },
\end{align}
\emph{b) and when $\gamma _{th}^{GF} < {{\gamma _{th}^{GB}} \mathord{\left/
 {\vphantom {{\gamma _{th}^{GB}} {\left( {\gamma _{th}^{GB} + 1} \right)}}} \right.
 \kern-\nulldelimiterspace} {\left( {\gamma _{th}^{GB} + 1} \right)}}$,}
\begin{align}
&P_{out,{p_2}}^{GB,II,\infty } = P_{out,{p_2}}^{GF,II,\infty } + {Q_4}\left( {{\sigma _3}} \right) + F_{{g_{GB}}}^{far}\left( {{\sigma _4}} \right) - F_{{g_{GB}}}^{far}\left( {{\sigma _3}} \right)\notag\\
 &\hspace*{0.2cm}- \frac{{{C_4}}}{{{\rho _{GF}}}}\left[ {M\left( {{\sigma _4},2 - {b_3},{b_3},b_{2,2}^{GB}} \right) - M\left( {{\sigma _4},2 - {b_3},{b_3},b_{2,1}^{GB}} \right)} \right.\notag\\
&\hspace*{0.2cm}\left. { - M\left( {{\sigma _3},2 - {b_3},{b_3},b_{2,2}^{GB}} \right) + M\left( {{\sigma _3},2 - {b_3},{b_3},b_{2,1}^{GB}} \right)} \right],
\end{align}
\emph{where ${Q_4}\left( x \right)$ is defined as }
\begin{align}
&{Q_4}\left( x \right) = F_{{g_{GB}}}^{far}\left( x \right) - {C_4}{\rho _{GB}}\notag\\
&\hspace*{0.3cm}\times \left[ {M\left( {x,2 - {b_3},{b_3},b_{2,2}^{GB}} \right) - M\left( {x,2 - {b_3},{b_3},b_{2,1}^{GB}} \right)} \right]\notag\\
&\hspace*{0.3cm} + {C_4}\left[ {M\left( {x,1 - {b_3},{b_3},b_{2,2}^{GB}} \right) - M\left( {x,1 - {b_3},{b_3},b_{2,1}^{GB}} \right)} \right].
\end{align}
\end{corollary}

\subsection{Asymptotic OP under Open-loop Protocol in Scenario II}

With the same assumptions of subsection~C, the asymptotic expressions under open-loop protocol are derived as \textbf{Corollary \ref{AII1}} and \textbf{Corollary \ref{asyfi}}.

\begin{corollary}\label{AII1}
\emph{Note that we assume the GF users experience high transmit power $P_{GF}$. Thus, we can achieve the asymptotic OP by the asymptotic expression of lower incomplete gamma functions as \eqref{gamma_asy}. With the assumption $P_{GF}  \to  \infty$, the asymptotic derivations for the GF users are derived.}

\emph{\textit{a)} Conditioned on ${\rho _{GF}} < {{\gamma _{th}^{GF}} \mathord{\left/
 {\vphantom {{\gamma _{th}^{GF}} {{\tau _{th}}}}} \right.
 \kern-\nulldelimiterspace} {{\tau _{th}}}}$, the asymptotic OP of the GF users can be derived as}
\begin{align}
P_{out,{p_1}}^{GF,II,\infty } = I_5^\infty  - F_{{g_{GF}}}^{near}\left( {{\tau _{th}}} \right).
\end{align}

\emph{\textit{b)} Conditioned on ${\rho _{GF}} > {{\gamma _{th}^{GF}} \mathord{\left/
 {\vphantom {{\gamma _{th}^{GF}} {{\tau _{th}}}}} \right.
 \kern-\nulldelimiterspace} {{\tau _{th}}}}$, we can derive the asymptotic OP expressions as}
 \begin{align}
&P_{out,{p_1}}^{GF,II,\infty } = \sum\limits_{s = 1}^S {{\Xi _{s,2}}\left[ {{C_3}{\sigma _2}\left( {\frac{{{\rho _{GB}}}}{{{\iota _{s,5}}}} + 1} \right) - F_{{g_{GF}}}^{near}\left( {{\tau _{th}}} \right)} \right]} \notag\\
 &\hspace*{0.3cm}\times \left[ {\gamma \left( {{b_3},b_{2,2}^{GB}\iota _{s,5}^{ - 1}} \right) - \gamma \left( {{b_3},b_{2,1}^{GB}\iota _{s,5}^{ - 1}} \right)} \right],
\end{align}
\emph{where ${\Xi _{s,2}} = \frac{1}{2}b_{1,2}^{GB}{\sigma _5}{\omega _s}\iota _{s,5}^{{b_3} - 2}{\left( {1 - \iota _{s,5}^2} \right)^{\frac{{\rm{1}}}{{\rm{2}}}}}$.}
\end{corollary}

\begin{corollary}\label{asyfi}
\emph{Utilizing the same assumption in \textbf{Corollary \ref{AII1}}, the asymptotic OP expressions of the GB users can be derived under the following three conditions.}

\emph{\textit{a)} When ${\rho _{GF}} < {{\gamma _{th}^{GF}} \mathord{\left/
 {\vphantom {{\gamma _{th}^{GF}} {{\tau _{th}}}}} \right.
 \kern-\nulldelimiterspace} {{\tau _{th}}}}$, $P_{out,{p_1}}^{GB,II,\infty }$ can be derived as}
\begin{align}
P_{out,{p_1}}^{GB,II,\infty } = {Q_4}\left( {{\sigma _2}} \right) + P_{out,{p_1}}^{GF,II,\infty }.
\end{align}
\emph{\textit{b)} When ${{\gamma _{th}^{GF}} \mathord{\left/
 {\vphantom {{\gamma _{th}^{GF}} {{\tau _{th}}}}} \right.
 \kern-\nulldelimiterspace} {{\tau _{th}}}} < {\rho _{GF}} < {{\gamma _{th}^{GF}\left( {1 + \gamma _{th}^{GB}} \right)} \mathord{\left/
 {\vphantom {{\gamma _{th}^{GF}\left( {1 + \gamma _{th}^{GB}} \right)} {{\tau _{th}}}}} \right.
 \kern-\nulldelimiterspace} {{\tau _{th}}}}$, the asymptotic expressions of OP can be calculated as}
\begin{align}
P_{out,{p_1}}^{GB,II,\infty } &= F_{{g_{GB}}}^{far}\left( {\sigma _5^{ - 1}} \right)\left[ {1 - F_{{g_{GF}}}^{near}\left( {{\tau _{th}}} \right)} \right] + {Q_4}\left( {{\sigma _2}} \right)\notag\\
 &\hspace*{0.3cm}- {Q_4}\left( {\sigma _5^{ - 1}} \right) + P_{out,{p_1}}^{GF,II,\infty }.
\end{align}

\emph{\textit{c)} When ${\rho _{GF}} \ge {{\gamma _{th}^{GF}\left( {1 + \gamma _{th}^{GB}} \right)} \mathord{\left/
 {\vphantom {{\gamma _{th}^{GF}\left( {1 + \gamma _{th}^{GB}} \right)} {{\tau _{th}}}}} \right.
 \kern-\nulldelimiterspace} {{\tau _{th}}}}$, we can derive the asymptotic expressions as}
\begin{align}
P_{out,{p_1}}^{GB,II,\infty } = \Theta \left[ {1 - F_{{g_{GF}}}^{near}\left( {{\tau _{th}}} \right)} \right] + P_{out,{p_1}}^{GF,II,\infty },
\end{align}
\emph{where $\Theta  = {{2{\sigma _2}\left( {R_2^{2 + \alpha } - R_1^{2 + \alpha }} \right)} \mathord{\left/
 {\vphantom {{2{\sigma _2}\left( {R_2^{2 + \alpha } - R_1^{2 + \alpha }} \right)} {\left[ {\left( {2 + \alpha } \right)\left( {R_2^2 - R_1^2} \right){\lambda _{GB}}} \right]}}} \right.
 \kern-\nulldelimiterspace} {\left[ {\left( {2 + \alpha } \right)\left( {R_2^2 - R_1^2} \right){\lambda _{GB}}} \right]}}$}.
\end{corollary}

\begin{remark}
Note that different protocols, i.e., open-loop protocol and dynamic protocol, have equivalent diversity gains. Compared to diversity orders in Scenario I and Scenario II, one conclusion can be obtained that constant diversity gains are obtained as 1) one for the near users and 2) zero for the far users.
\end{remark}
\section{Numerical Results}
In this section, numerical results are indicated to validate analytical, approximated and asymptotic expressions derived in the previous sections, and further facilitate the outage performance and analysis of diversity orders.
\begin{table}[!t]
\centering
\caption{Diversity orders for the GB and GF users under two scenarios with different SIC orders.}

\linespread{1}

\label{table}
\begin{center}
\begin{tabular}{ | m{2cm} | m{2cm}| m{2cm} | }
\hline
Diversity orders & The GF Users & The GB users  \\
\hline
Scenario I &  0 & 1  \\
\hline
Scenario II &  1 &  0  \\
\hline
\end{tabular}
\end{center}

\end{table}

\subsection{Simulation Results on Outage Performance in Scenario I}
\begin{figure}[!htb]
\centering
\includegraphics[width= 3.6in]{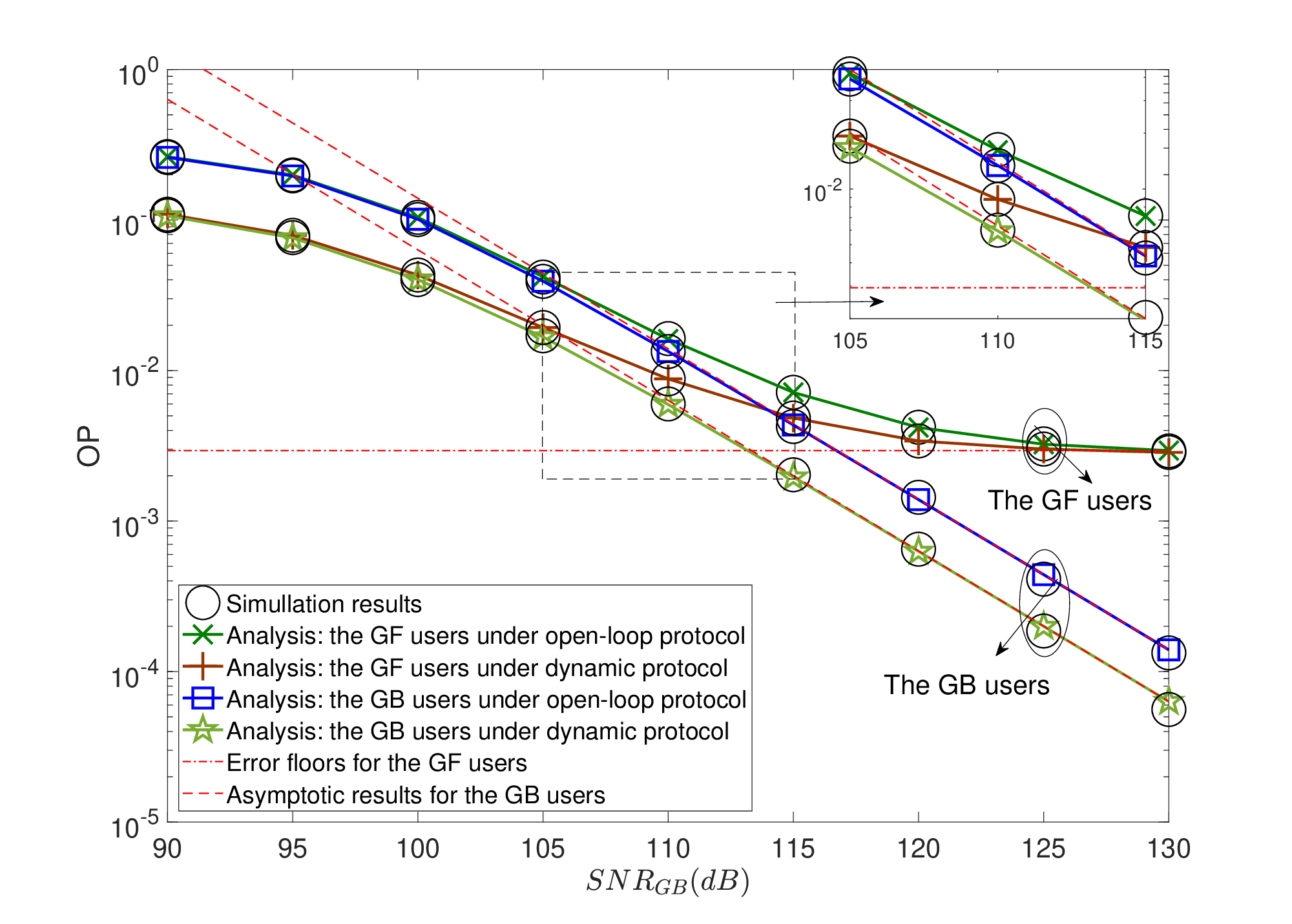}
\caption{Outage probability v.s. transmit SNR for the GB users $\rho_{GB}=[90,130]$ dB with variations of protocols - Scenario I}
\label{1}
\end{figure}
\begin{figure}[!htb]
\centering
\includegraphics[width= 3.6in]{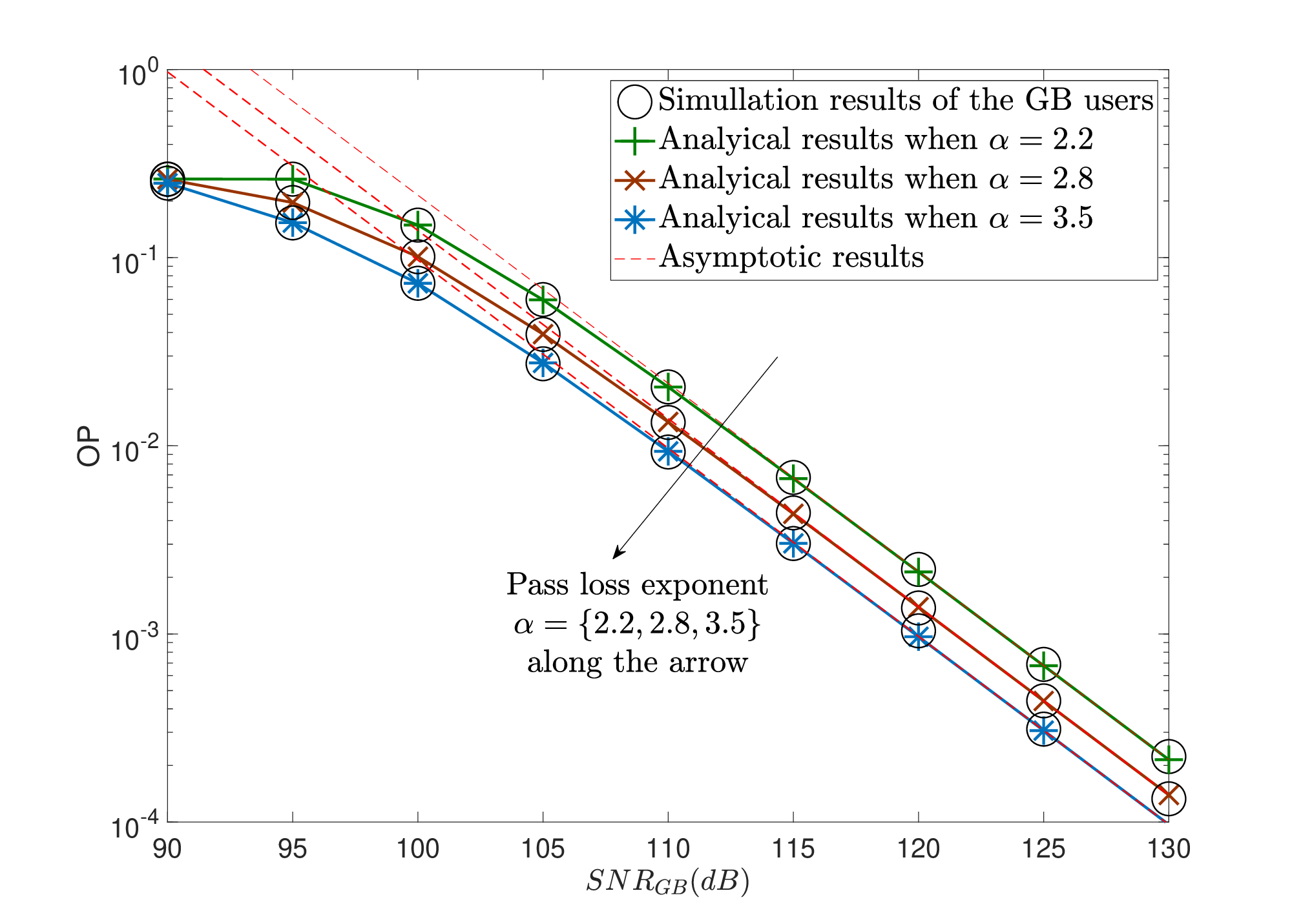}
\caption{Outage probability of the GB users v.s. transmit SNR of the GB users $\rho_{GB}=[90,130]$ dB with variations of pass loss exponent $\alpha=[2.2,2.8,3.5]$ under open-loop protocol - Scenario I}
\label{2}
\end{figure}
\begin{figure}[!htb]
\centering
\includegraphics[width= 3.6in]{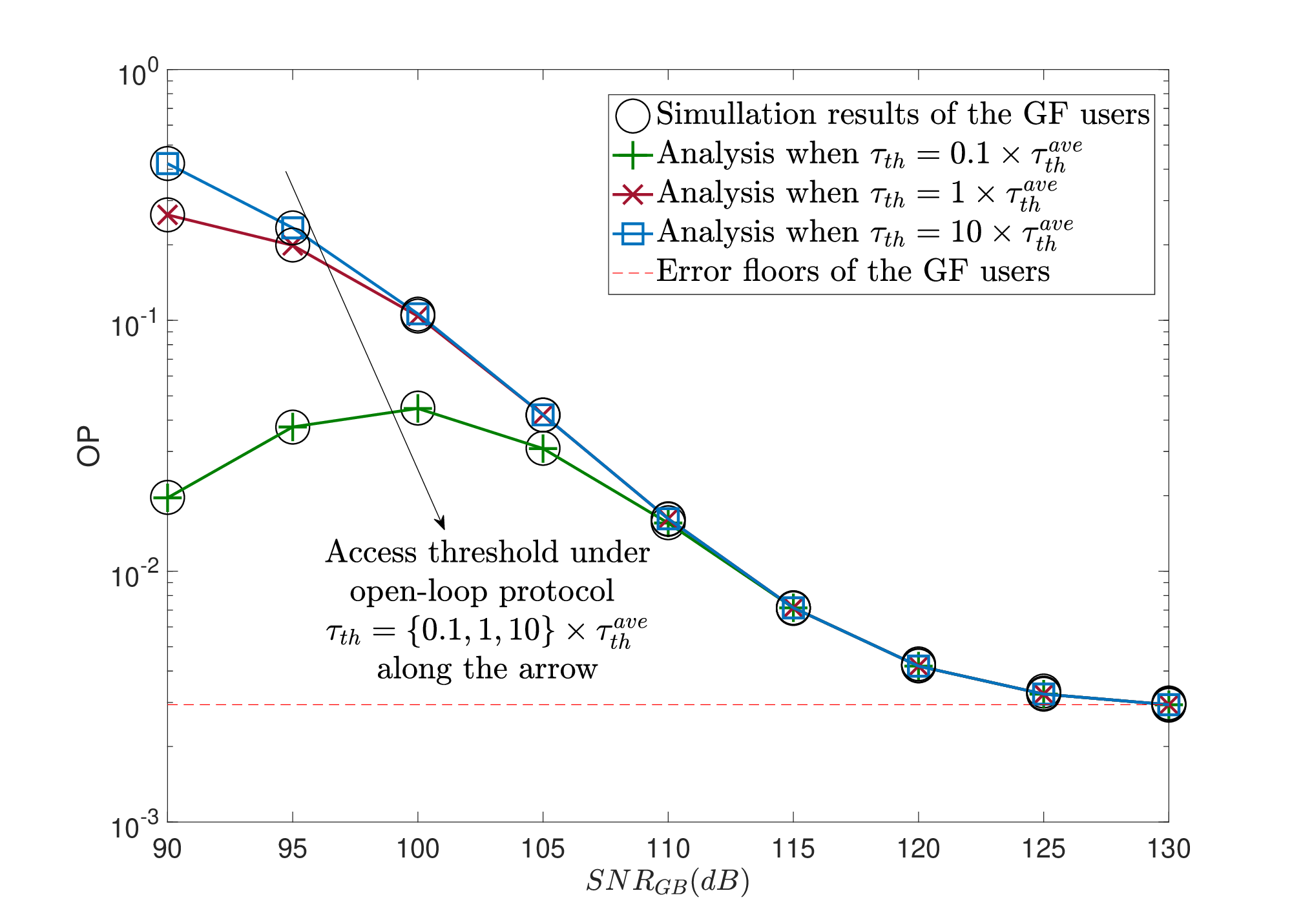}
\caption{Outage probability of the GF users v.s. transmit SNR for the GB users $\rho_{GB}=[90,130]$ dB with variations of access threshold $\tau_{th}$ under open-loop protocol - Scenario I}
\label{3}
\end{figure}

\subsubsection{Validation of Results - Scenario I}
We first validate the analytical results and investigate the impact of distances on outage probability. In this subsection, coefficients are fixed unless otherwise specified. Note that the BS is located at the center of the disc, the locations of devices are drawn from uniformly distributed circle region with the radium of the disc as $R_1 = 200$ m and the radius of the ring as $(R_1,R_2) = (200, 600)$ m. Other coefficients set as follows: the channel gain for the GB and the GF users ${\left| {{h_{GB}}} \right|^2}$ and ${\left| {{h_{GF}}} \right|^2}$ as $0$ dB, pass-loss exponent $\alpha$ as 2.8, outage thresholds for the GB and GF users as $R_{GF}=1$ BPCU and $R_{GB}=1.5$ BPCU, where BPCU means bit per channel use, noise power $\sigma^2$ as $-90$ dBM, which is calculated as $\sigma^2 = 170 + 10\log_{10}(BW)+N_f$, where carrier frequency $BW$ is $10$ MHz, the noise figure $N_f=10$ dB. Additionally, we set channel quality thresholds of open-loop protocol $\tau_{th}$ as the mean of $P_{GB}P_{GF}^{-1}{\left| {{h_{GB}}} \right|^2}{\left( {{d_{GB}}} \right)^{ - \alpha }}$. In Scenario I, the transmit power of the GF users are fixed as $10$ dBM and that of the GB users varies from $[0,40]$ dBM. Comparing to the simulation and analytical results from Fig. \ref{1} to Fig. \ref{3}, all curves are perfect matches, thereby validating our analysis of four theorems from \textbf{Theorem \ref{OPGB2}} to \textbf{Theorem \ref{OPGF1}}. We also note that asymptotic expressions match the analytical ones in high SNR region, which verifies the accuracy for our asymptotic analysis.

\subsubsection{Impact of Protocols on Outage Probability - Scenario I}

In Scenario I, the outage performance of the GF and GB users is investigated under two protocols indicated in Fig. \ref{1}. One common observation on simulation results indicates that the dynamic protocol outperforms open-loop protocol for all users. This is because under dynamic protocol, frequent transmissions of the threshold by an added handshake can maintain the accuracy of access thresholds when the locations of the GB users are changed. However, under open-loop protocol, the BS transmits an average threshold to all the GF users, which may cause more interference.

\subsubsection{Impact of Pass Loss on Outage Probability - Scenario I}
In Fig. \ref{2}, we investigate the influence of various pass loss on OP. Numerical results demonstrate that the users with a large pass loss exponent present high OP. The reason is that the high pass loss exponent can reduce received power of all users but the far users fade severely, thereby the interference from the GF users to the GB users can be reduced by enhancing the value of pass loss exponent in an appropriate range.

\subsubsection{Impact of Access threshold on Outage Probability - Scenario I}
We set the access threshold $\tau_{th}$ as the mean of $P_{GB}P_{GF}^{-1}{\left| {{h_{GB}}} \right|^2}{\left( {{d_{GB}}} \right)^{ - \alpha }}$, denoted as $\tau_{th}^{ave}$, because the threshold should be correlated to the channel gain of the GB users $g_{GB}$ to meet the access conditions as $P_{GF}g_{GF}<P_{GB}g_{GB}$ in Scenario I or $P_{GF}g_{GF}>P_{GB}g_{GB}$ in Scenario II. Thus, in Fig. \ref{3}, the effect on OP of the GF users caused by access threshold $\tau_{th}$ is studied via one-tenth, one or ten times of $\tau_{th}^{ave}$. we observe that a low threshold can achieve superior performance because more GF users can be accepted into the channel to transmit messages with a small access threshold.

\subsection{Simulation Results on Outage Performance in Scenario II}
\begin{figure}[!htb]
\centering
\includegraphics[width= 3.6in]{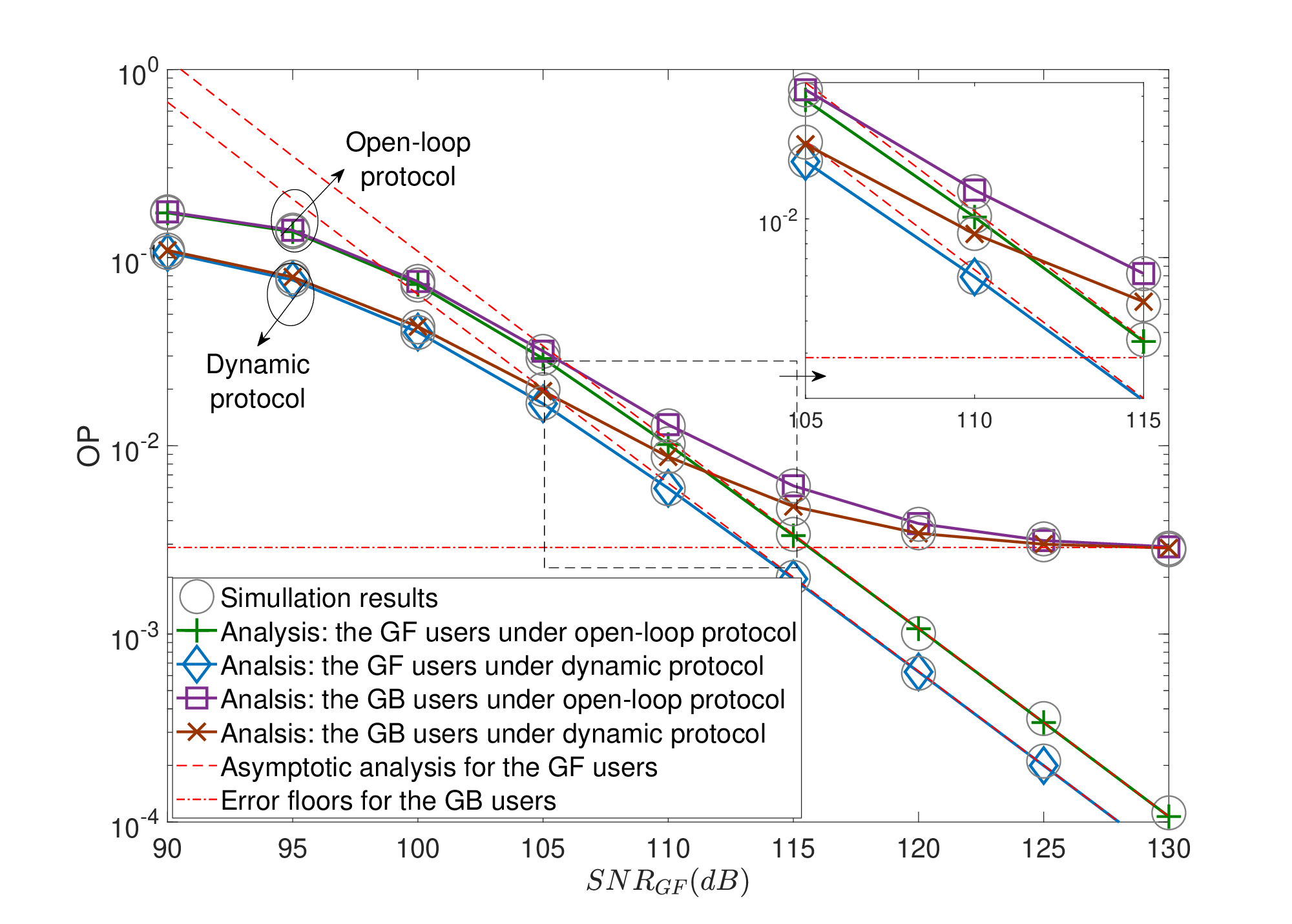}
\caption{Outage probability v.s. transmit SNR for the GF users $\rho_{GF}=[90,130]$ dB with different protocols - Scenario II}
\label{4}
\end{figure}
\begin{figure}[!htb]
\centering
\includegraphics[width= 3.6in]{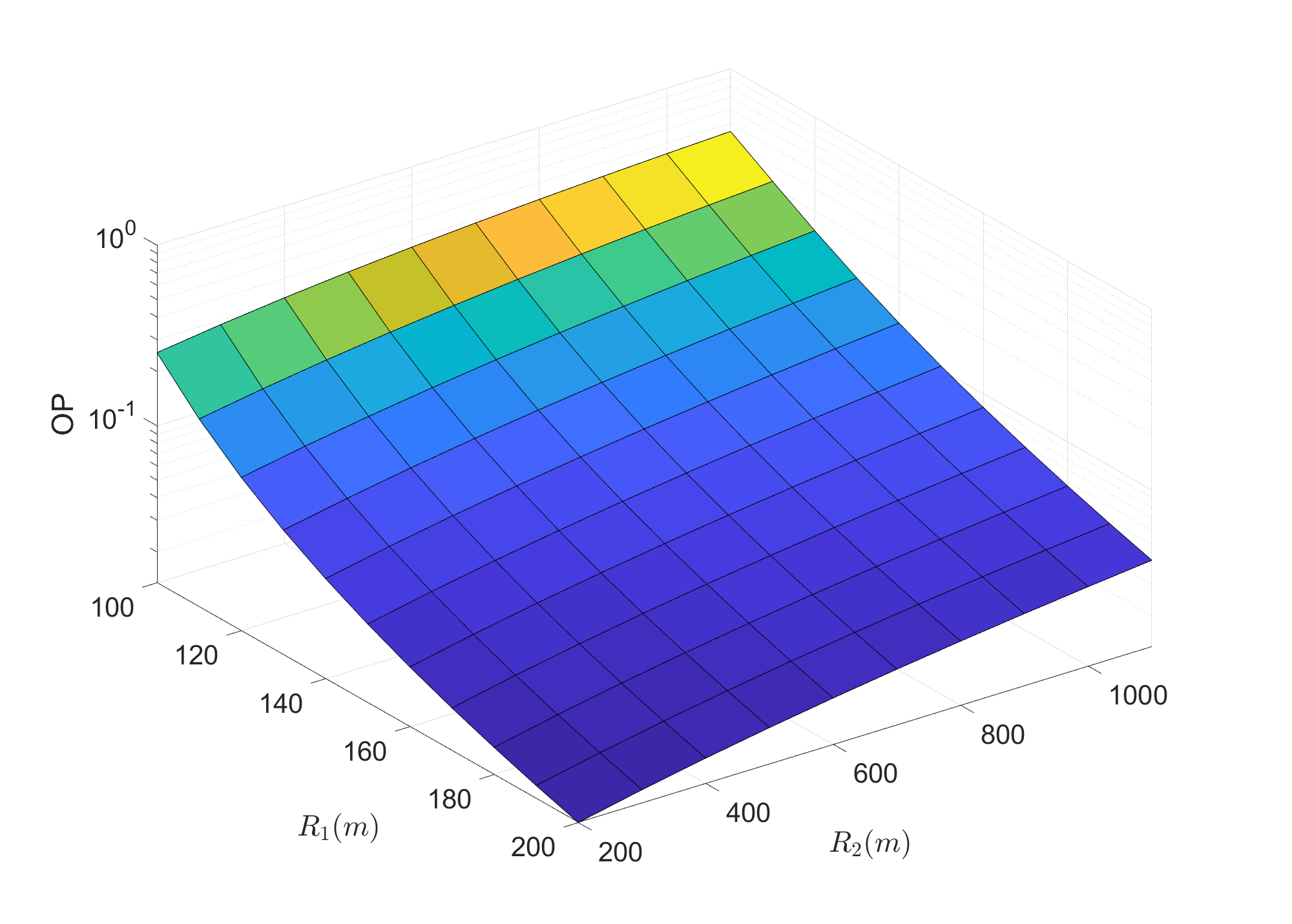}
\caption{Outage probability of the GF users v.s. the radius of the ring and disc $(R_1,R_2)$ under open-loop protocol - Scenario II}
\label{5}
\end{figure}
\begin{figure}[!htb]
\centering
\includegraphics[width= 3.6in]{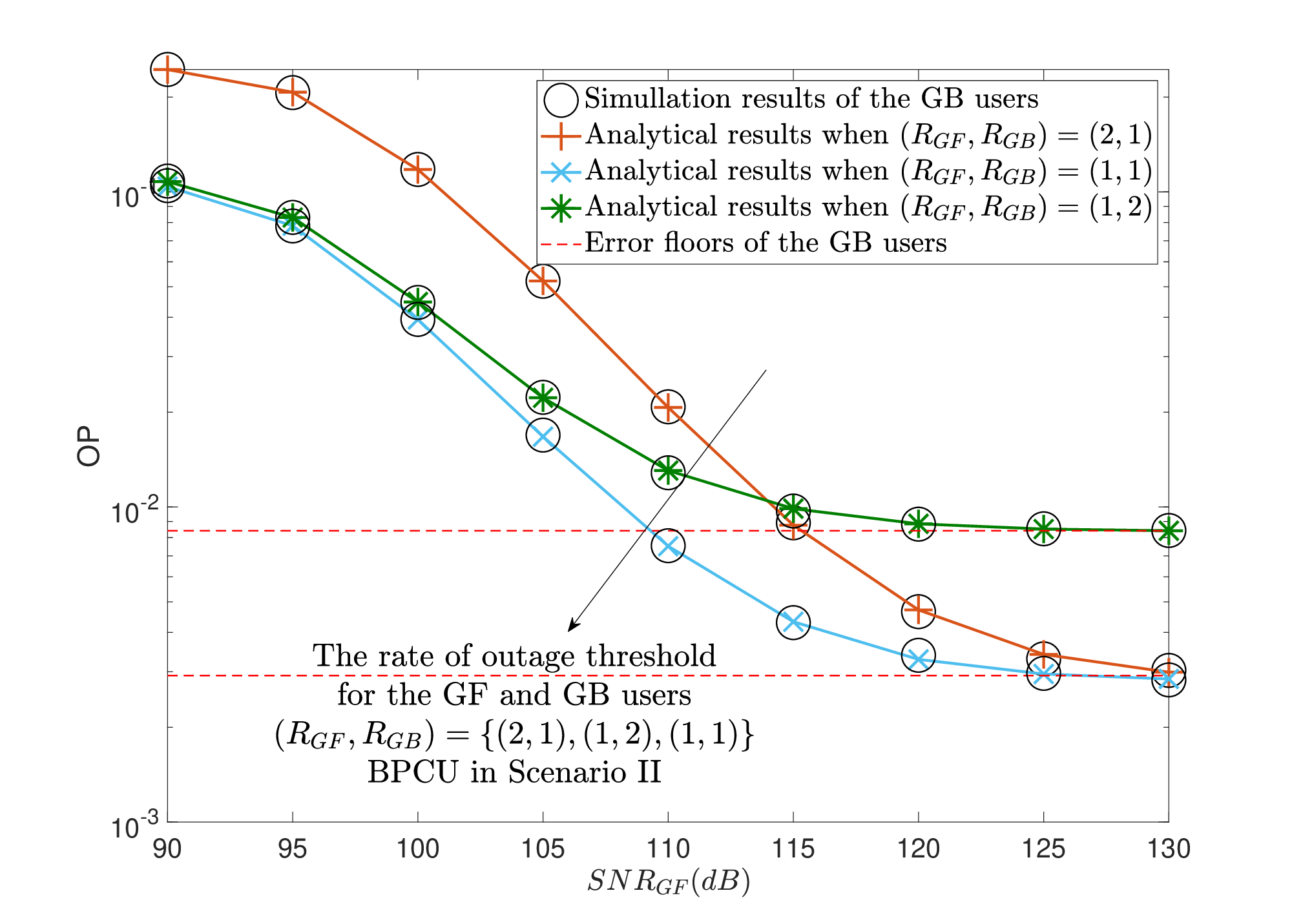}
\caption{Outage probability of the GB user v.s. transmit SNR for the GF users $\rho_{GF}=[90,130]$ dB with variations of outage thresholds $(R_{GB},R_{GF})$ - Scenario II}
\label{6}
\end{figure}
\subsubsection{Validation of Results - Scenario II}
Validation of analytical results on OP in Scenario II is demonstrated as a perfect match by Fig. \ref{4} to Fig. \ref{6}, including four theorems, i.e., from \textbf{Theorem \ref{GF_p2_T2}} to \textbf{Theorem \ref{GB_p1_T2}}. Additionally, curves for asymptotic expressions match simulation results in high SNR region to verify \textbf{Corollary \ref{asy1}} and \textbf{Corollary \ref{asyfi}}. Without otherwise specification, we set the same numerical coefficients of outage performance in Scenario I except for the transmit power for the GF users as $[0,40]$ dBM and the transmit power of the GB users as $10$ dBM. Thus, the OP can be indicated by Fig. \ref{4} versus the transmit SNR of the GF users. In Fig. \ref{1} and Fig. \ref{4}, one conclusion can be summarized that near users obtain better diversity gains than far users.

\subsubsection{Impact of Distance on Outage Probability - Scenario II}
In Fig. \ref{5}, analytical results of OP v.s. the radius of the disc and ring $(R_1,R_2)$ are depicted for the GF users. One can be observed that the outage probability decreases as the radium of the disc $R_1$ increases while inclines when the outer radium of the ring $R_2$ improves. This is because we invoke stochastic geometry to present spatial effect of user locations, thereby increasing $R_1$ improves the extent of large-scale fading with higher values of outage probability for the GF users. Additionally, improving $R_2$ means reducing the interference from the GB users, which causes better channel conditions.

\subsubsection{Impact of Outage Threshold on Outage Probability - Scenario II}

The last figure illustrates the effect of the outage threshold $R_{GF}$ and $R_{GB}$ on OP. In Fig. \ref{6}, OP of the GB users versus the transmit SNR of the GF users is analyzed via various combinations of the rates of the threshold $R_{GF}$ and $R_{GB}$. It can be concluded that declining the threshold of the GF users can enlarge the performance of the GB users. This is because a low threshold means satisfying channel condition. Another can be obtained that reducing the threshold of the GB users can enhance the error floors of the GB users. Note that this figure investigates the performance of the GB users, thus $R_{GB}$ determines the error floors.

\section{Conclusions}

Uplink semi-GF NOMA networks have been investigated to obtain reduced collision situations and enhanced spectrum efficiency. Stochastic geometry has been invoked to capture the spatial effects of NOMA users. We propose a novel contention control protocol, denoted as dynamic protocol, to select which portion of the GF users are employed into NOMA transmissions. We utilize the open-loop protocol as the benchmark. Compared with open-loop protocol, dynamic protocol provides more accurate channel quality thresholds, which enables to reduce the interference from the GF users. As the locations of the GF and GB users are not clarified, two potential scenarios to determine the SIC orders have been proposed that: 1) the GB users as near users are decoded firstly in Scenario I and 2) the GF users are near users in Scenario II. Based on the two scenarios, outage probabilities have been derived via analytical, asymptotic and approximated expressions for the GB and GF users. Analytical results have concluded that under two scenarios for both protocols, consistent diversity gains are determined by the SIC orders that equal to 1) one for near users and 2) zero for far users. Validated by numerical results, we reveal that dynamic protocol enhances the outage performance than open-loop protocol.

\numberwithin{equation}{section}


\end{document}